\documentclass[reprint,aps,prb]{revtex4-1}
\usepackage{amsmath,amssymb}
\usepackage{graphicx}
\usepackage{color}
\usepackage{bm}

\newcommand{\Part}[3]{ \frac{ \partial^{#3} #1 }{ \partial #2^{#3} } }
\newcommand{\V}[1]{\bm{#1} } 

\newcommand{\bra}[1]{\left \langle {#1} \right| } 
\newcommand{\ket}[1]{\left| {#1} \right\rangle } 

\newcommand{\mN}{\mathbb{N}}
\newcommand{\mZ}{\mathbb{Z}}
\newcommand{\lb}{\left(}
\newcommand{\rb}{\right)}
\newcommand{\lbb}{\left\{}
\newcommand{\rbb}{\right\}}

\newcommand{\Blbb}{ \Biggl\{ }
\newcommand{\Brbb}{ \Biggr\} }

\newcommand{\T}[1]{\tilde{#1}}

\newcommand{\BReq}[1]{Eq.~(\ref{eq:#1})}
\newcommand{\NReq}[1]{(\ref{eq:#1})}

\newcommand{\Rfig}[1]{Fig.~\ref{fig:#1}}

\newcommand{\NRfig}[1]{\ref{fig:#1}}
\newcommand{\Lfig}[1]{\label{fig:#1}}
\newcommand{\Leq}[1]{\label{eq:#1}}
\newcommand{\Rsec}[1]{Sec.~\ref{sec:#1}}
\newcommand{\Lsec}[1]{\label{sec:#1}}
\newcommand{\be}{\begin{eqnarray}}
\newcommand{\ee}{\end{eqnarray}}
\newcommand{\ba}{\begin{array}}
\newcommand{\ea}{\end{array}}
\newcommand{\no}{\nonumber}

\newcommand{\subbe}{\begin{subequations}}
\newcommand{\subee}{\end{subequations}}

\newcommand{\argmin}{\mathop{\rm arg~min}\limits}

\newcommand{\ketup}{\ket{\uparrow}}
\newcommand{\ketdown}{\ket{\downarrow}}
\newcommand{\braup}{\bra{\uparrow}}
\newcommand{\bradown}{\bra{\downarrow}}

\begin{document} 

\title{Complex semiclassical analysis of the Loschmidt amplitude and \\ 
dynamical quantum phase transitions}
\author{Tomoyuki Obuchi$^{1}$}\email{obuchi@c.titech.ac.jp}
\author{Sei Suzuki$^{2}$}
\author{Kazutaka Takahashi$^{3}$}
\affiliation{
$^{1}$Department of Mathematical and Computing Science, 
Tokyo Institute of Technology, Yokohama 226-8502, Japan
\\
$^{2}$Department of Liberal Arts, Saitama Medical University, 
Moroyama, Saitama 350-0495, Japan
\\
$^{3}$Department of Physics, Tokyo Institute of Technology, Tokyo 152-8551, Japan
}
\date{\today}

\begin{abstract}
We propose a new computational method of 
the Loschmidt amplitude in a generic spin system on the basis of 
the complex semiclassical analysis on the spin-coherent state path integral.
We demonstrate how the dynamical transitions emerge in
the time evolution of the Loschmidt amplitude for the infinite-range transverse Ising model with a longitudinal field, exposed by a quantum quench of the transverse field $\Gamma$ from $\infty$ or $0$.
For both initial conditions, we obtain the dynamical phase diagrams
that show the presence or absence of the dynamical transition 
in the plane of transverse field 
after a quantum quench and the longitudinal field.
The results of semiclassical analysis are verified by numerical experiments.
Experimental observation of our findings on the dynamical transition is 
also discussed.
\end{abstract}

\maketitle 

\section{Introduction}
Triggered by experiments using ultracold atomic systems, dynamics of a
closed quantum many-body system has been one of the fascinating topics
in condensed matter physics~\cite{Polkovnikov:11}.
In particular, the time evolution after a sudden change of the Hamiltonian
has attracted a lot of attention as a basic setting
of a problem on the out-of-equilibrium quantum state. 
One of the interesting phenomenon associated with this so-called 
quantum quench is the dynamical quantum phase transition (DQPT). 
While the equilibrium quantum phase transition
is usually associated with a singularity of the ground-state energy
in the axis of a parameter contained in the Hamiltonian,
the DQPT involves a singularity in time. 
The present paper focuses on such a dynamical singularity
appearing in the return probability to the initial state,
which is directly related to the Loschmidt amplitude defined below~\cite{Heyl:13}.

The phenomena of the DQPT are observed not only 
in the Loschmidt amplitude but also in the time average of local physical
quantities such as order parameters. 
Although a certain correspondence is pointed out~\cite{Zunkovic:16}, 
these two kinds of quantities are generally different.
The local physical quantities represent the properties of the steady state in
the long time limit after a quantum quench. 
They bring a clear physical consequence and are easy to access by experiments.
The DQPT of them corresponds to a phase transition with the parameter
in the Hamiltonian after a quantum quench.
The Loschmidt amplitude, on the other hand,
can be seen as an extension 
of the partition function on the imaginary axis corresponding to time.
The DQPT here is defined as a singular behavior
with time in the rate function of it, as an analogy
with the thermodynamic phase transition accompanied by the singularity
of the free energy as a function of the temperature.
However, the Loschmidt amplitude
involves delicate points 
in several aspects: physical meaning of the singularities,
experimental implementations, and even technicalities for theoretical computations.
Several recent works~\cite{Heyl:13,Heyl:14,Heyl:15,Zunkovic:15,Zunkovic:16,
Karrasch:13,Vajna:14,Sharma:15,Divakaran:16,Gambassi:11-1,Gambassi:11-2,Chiocchetta:15,Chiocchetta:16,Maraga:15,Smacchia:13,Gambassi:12}
have been devoted to resolve the first delicate point based on
statistical mechanical concepts such as renormalization group, symmetry
breaking, universality, and scaling~\cite{NishimoriBook}. 
They have provided solid advances.
For instance, the singularity has been tied with a behavior of the order parameter and entanglement production in systems with symmetry-broken phases~\cite{Heyl:14,Jurcevic:16}. However, a general comprehension including
the relation of the singularities to other local quantities 
with a generic initial state is still lacking.
One of the origins of the difficulty in obtaining a general
description lies, in our opinion, 
in the limitation on theoretical techniques
to compute the Loschmidt amplitude. 
Most of theoretical works so far depend on the result of specific models 
being analytically tractable, and generic properties of the Loschmidt 
amplitude's singularity have been speculated from the result. 
Hence, a more versatile computational method will be a great help 
to understand the Loschmidt amplitude.  

Under this circumstance, here we propose a new theoretical framework for
computing the Loschmidt amplitude for a generic spin system on the basis of 
a semiclassical computation. 
This can be regarded as a mean-field method and is expected to be exact 
in the infinite dimension, though it is still applicable 
as an approximation to a generic spin system in any
dimension with an arbitrary state.
The static approximation is often used with the
mean-field method and is known to give a correct result for quantities
in the equilibrium in the system with an infinite-range interaction \cite{SuzukiBook}. 
However, the static approximation does not work for 
the computation of the Loschmidt amplitude.
In this sense, our method goes beyond the static approximation and
can be useful for computation of out-of-equilibrium quantities.

Our semiclassical method is essentially the same as the one used
in Refs.~\cite{Zunkovic:15,Sciolla:10,Sciolla:11}, but their analysis has been
only on local physical quantities. 
This is presumably due to the lack of general prescriptions
to compute the Loschmidt amplitude so far. 
The present work complements this point.  
The key difference of our method from the preceding
studies~\cite{Zunkovic:15,Sciolla:10,Sciolla:11}
lies in the determination of the semiclassical path
that follows the initial and final conditions properly.
In our method, the range of
dynamical variables is extended from real to complex numbers, 
and matching the semiclassical path with 
the boundary conditions is achieved in the complex space. 
This idea has been proposed in Refs.~\cite{Klauder:79,Alscher:99} 
for single-spin systems and we extensionally apply it to many-spin systems. 
Accordingly, there emerge multiple solutions in the boundary value problem, and
the solution that gives the largest return probability is selected.
We find that it is this selection that
gives the singularity in the Loschmidt amplitude.
In this sense, the singularity of Loschmidt amplitude
is very similar to an equilibrium phase transition.

Using the complex semiclassical method, in the present paper, we study the
infinite-range transverse Ising model Hamiltonian 
with uniform coupling $J$ and longitudinal field $h$:
\be
 \hat{\mathcal{H}}=
 -\frac{NJ}{2}\lb \frac{1}{N}\sum_{i=1}^{N}\sigma_i^{z} \rb^2
 -\Gamma \sum_{i=1}^{N}\sigma_i^{x}-h\sum_{i=1}^{N}\sigma_i^{z},
 \Leq{Hamiltonian}
\ee
where $\sigma_i^{\alpha}$
($i=1,2,\dots,N$; $\alpha=x,z$) is the Pauli matrix and
$N$ is the number of spins.
For this model, we consider a quantum quench of the transverse field $\Gamma$
from $\Gamma_{\rm i}$ to $\Gamma_{\rm f}$ at $t = 0$. 
As shown in \Rfig{PD_static}, this system
shows two different phases in equilibrium \cite{SuzukiBook} and 
both inter- and intra-phase protocols of quench are examined.
\begin{figure}[tbp]
\begin{center}
\includegraphics[width=0.53\columnwidth]{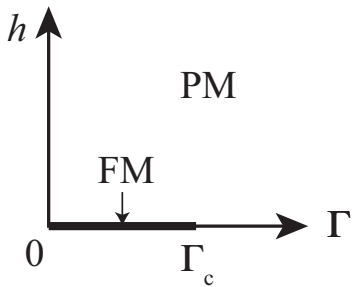}
\caption{Phase diagram of the ground state in the infinite-range
transverse Ising model \NReq{Hamiltonian} in the plane of the
transverse field $\Gamma$ and the symmetry-breaking longitudinal field $h$. 
The ferromagnetic (FM) phase lies on the axis of $\Gamma$ from $0$ to
$\Gamma_{\rm c}=J$ with $h = 0$,
while other parameter area is a paramagnetic (PM) phase.}
\Lfig{PD_static}
\end{center} 
\end{figure}
The Loschmidt amplitude is defined by
\be
 \mathcal{L}\lb t |\psi \rb=\bra{\psi} e^{-it\hat{\mathcal{H}}} \ket{\psi},
\ee
where the state $\ket{\psi}$ is chosen as the ground state 
of the Hamiltonian with $\Gamma = \Gamma_{\rm i}$. 
The Loschmidt amplitude is
expected to exhibit a large deviation nature, and hence its rate
function at $N\to\infty$ is the primary object of our analysis. The
  rate function is defined as
\be
 f\lb t| \psi \rb =-\frac{1}{N}\log \mathcal{L}\lb t |\psi \rb.
 \label{ratef}
\ee
Note that its real part, $f_{\rm r}=\Re{f}$, accounts for the return 
probability $P(t|\psi)=|\mathcal{L}\lb t |\psi \rb|^2$ 
as $2f_{\rm r}=-\frac{1}{N}\log P(t|\psi)$, while the imaginary part 
has no direct physical consequence. 

The rest of the paper is organized as follows. 
In Sec.~\ref{Formulation}, we describe the formulation and 
procedures needed to make the problem computationally tractable. 
In Sec.~\ref{Result}, the analytical solutions
computed from the invented method are shown and are compared to
numerical experiments on finite size systems. 
Exact derivation of the rate function, 
available only on some specific parameters, 
is also given to justify the result. 
Section \ref{Discussion} is devoted to discussion and summary. 
The relevance of the present work to experiments, quantum engineering, 
and computation is discussed there.

\section{Formulation}
\label{Formulation}

\subsection{Spin coherent states and path integrals}
\Lsec{Spin coherent}

We start from reviewing the path integral formulation for spin systems.
An arbitrary state of a single spin is represented by 
a spin-coherent state as
\be
 \ket{\theta,\varphi}
 =e^{ib}\lb e^{-i\frac{\varphi}{2}}\cos{\frac{\theta}{2}}\ketup+e^{i\frac{\varphi}{2}}
 \sin{\frac{\theta}{2}}\ketdown \rb,
\ee
where $\ketup$ and $\ketdown$ are the eigenstates of $\sigma^z$ 
with eigenvalues $+1$ and $-1$, respectively.
Hereafter the gauge $b$ is fixed to be $0$ and is disregarded,
since it does not affect any physical consequences.
As is well known, the average of spin variables 
$\V{\sigma}=(\sigma^x,\sigma^y,\sigma^z)$  
over a spin-coherent state corresponds to three-dimensional polar representation as 
\be
 \bra{\theta,\varphi}\V{\sigma}\ket{\theta,\varphi}
 = \lb \sin{\theta}\cos{\varphi},\sin{\theta}\sin{\varphi},\cos{\theta}\rb.
 \Leq{polar}
\ee
The spin-coherent state constitutes an overcomplete basis:
\be
 \int_{-1}^1 d\cos\theta\int_{0}^{2\pi}\frac{d\varphi}{2\pi}
 \ket{\theta,\varphi}\bra{\theta,\varphi}=
 \ketup\braup +\ketdown\bradown=I, \nonumber\\
 \Leq{complete}
\ee
where $I$ denotes the $2\times 2$ unit matrix. 
Note that two states with different $(\theta,\varphi)$ are not orthogonal in general.

We apply this spin-coherent state formulation to
$N$-spin systems and write the variables as
$(\V{\theta},\V{\varphi})=\{(\theta_{i},\varphi_i)\}_{i=1}^{N}$. 
Using the spin-coherent states, we write any propagator with arbitrary time-dependent
Hamiltonian $\hat{\mathcal{H}}(t)$ as $G(t|\Omega',\Omega'')\equiv
\bra{\Omega''}\mathcal{T}e^{-i\int^{t}_0 ds
\hat{\mathcal{H}}(s)}\ket{\Omega'}$, where $\mathcal{T}$ is the time-ordering
operator, and $\Omega'=(\V{\theta}',\V{\varphi}')$ and
$\Omega''=(\V{\theta}'',\V{\varphi}'')$ are initial and final 
states respectively. 
This propagator is rewritten in a path integral form as~\cite{Klauder:79} 
\be
 G(t|\Omega',\Omega'')
 =\int_{\Omega'}^{\Omega''} \prod_{i=1}^{N}\mathcal{D}\cos{\theta_i}
 \mathcal{D}\varphi_i~e^{S[\V{\theta},\V{\varphi}]}.
 \Leq{pathintegral}
\ee
This is an integral over all possible paths 
of the variables $(\V{\theta}(s),\V{\varphi}(s))$.  
The action functional $S[\V{\theta},\V{\varphi}]$ is given by 
\be
 S[\V{\theta},\V{\varphi}]=i\int_{0}^{t} ds
 \lbb\frac{1}{2}\sum_{i}\dot{\varphi}_i(s)\cos{\theta_i(s)}-\mathcal{H}
 \lb \V{\theta},\V{\varphi},s\rb\rbb,  \nonumber\\
\ee
where the dot symbol denotes the time derivative and 
$\mathcal{H}\lb \V{\theta},\V{\varphi},s\rb
=\langle \V{\theta}(s),\V{\varphi}(s)|\hat{\mathcal{H}}
|\V{\theta}(s),\V{\varphi}(s)\rangle$.

\subsection{Complex semiclassical analysis}
\Lsec{Extended semiclassical}

The path integral formalism gives the exact result 
if we can perform the integration over all paths literally. 
However, this is difficult in general, 
and the semiclassical approximation is here employed. 

\subsubsection{Boundary value problem}
\Lsec{Naive boundary condition}

The basic idea of the semiclassical method is to take into account 
only the dominant stationary paths among all the paths. 
The stationary condition in the action $S$ leads to 
the following equations of motion (EOMs):
\be
 \frac{1}{2}\dot{\theta}_i\sin \theta_i =\Part{\mathcal{H}}{\varphi_i}{},\qquad
 \frac{1}{2}\dot{\varphi}_i\sin \theta_i =-\Part{\mathcal{H}}{\theta_i}{}.
 \Leq{EOM-theta}
\ee
We naively expect that the solution of these EOMs, satisfying
the boundary conditions $\lb \V{\theta}(0),\V{\varphi}(0)\rb=\Omega'=\lb
\V{\theta}',\V{\varphi}'\rb$ and 
$\lb\V{\theta}(t),\V{\varphi}(t)\rb=\Omega''=\lb\V{\theta}'',\V{\varphi}''\rb$, 
is the desired semiclassical path. 
If there are multiple semiclassical paths, we give indices to them as 
$\{(\bar{\V{\theta}}^{(\nu)},\bar{\V{\varphi}}^{(\nu)})\}_{\nu}$ 
where the symbol $\bar{\cdot}$ represents a generic semiclassical path hereinafter. 
Semiclassical actions corresponding to those paths are defined as 
$S_{\rm cl}[\bar{\V{\theta}}^{(\nu)},\bar{\V{\varphi}}^{(\nu)}]=
S[\bar{\V{\theta}}^{(\nu)},\bar{\V{\varphi}}^{(\nu)}]$. 
They give an approximation of the propagator as  
\be
 G(t|\Omega',\Omega'') \sim  
 \sum_{\nu} A_\nu e^{S_{\rm cl}[\bar{\V{\theta}}^{(\nu)},\bar{\V{\varphi}}^{(\nu)}]},
 \Leq{G-naive}
\ee
where $A_\nu$ denotes a possible amplitude factor.

Unfortunately, this procedure does not work in the present problem. 
The solution of \BReq{EOM-theta} cannot satisfy, in general, both
the boundary conditions $\lb \V{\theta}(0),\V{\varphi}(0)\rb=\Omega'$
and $\lb \V{\theta}(t),\V{\varphi}(t)\rb=\Omega''$. 
For a given initial condition $\lb \V{\theta}(0),\V{\varphi}(0)\rb=\Omega'$, 
the time evolution of the system is uniquely determined by the EOMs,
and the final values $\lb \V{\theta}(t),\V{\varphi}(t)\rb$ do not
necessarily coincide with the boundary one $\Omega''$. 
This is the reason why the Loschmidt amplitude has been
difficult to be evaluated by the semiclassical or similar methods,
though some exceptions are found when the semiclassical path is constant
in time~\cite{Zunkovic:15,Obuchi:12,Takahashi:13}. 
To overcome this problem, 
following the prescription in Refs.~\cite{Klauder:79,Alscher:99}, 
we below introduce the so-called Wiener regularization term making the path integral 
well defined in the action, and deal with the unregularized action 
as the vanishing limit of the regularization term. 
This yields a different boundary condition. 

\subsubsection{Wiener regularization and modified boundary condition}
\Lsec{Wiener regularization and}

By using the prescription by Klauder~\cite{Klauder:79}, 
Alscher and Grabert demonstrated that the exact propagator
can be computed in single-spin systems 
with arbitrary time-dependent magnetic fields~\cite{Alscher:99}. 
Here we apply this to many-spin systems.  

The Wiener regularization is defined as
\be
 W[\V{\theta},\V{\varphi}] 
 = -\frac{1}{4}m\int_{0}^{t} ds\sum_{i}\lb \dot{\theta}_i^2
 + \dot{\varphi}_i^2\sin^2\theta_i \rb,
\ee
where $m$ represents a constant.
Adding this term to the action, 
$S[\V{\theta},\V{\varphi}]\to S[\V{\theta},\V{\varphi}]+W[\V{\theta},\V{\varphi}]$, 
and taking the stationary condition, we obtain the modified semiclassical EOMs as
\subbe
 \Leq{EOM-theta-regularized}
\be
 && \frac{1}{2}\dot{\theta}_j\sin \theta_j \nonumber\\
 && =\Part{\mathcal{H}}{\varphi_j}{} 
 +\frac{i}{2}m\left( \ddot{\varphi}_j \sin^2 \theta_j 
 +2\dot{\theta}_j\dot{\varphi}_j\sin{\theta_j}\cos{\theta_j} \right), \\ 
 && \frac{1}{2}\dot{\varphi}_j\sin\theta_j \nonumber \\
 && =-\Part{\mathcal{H}}{\theta_j}{} 
 -\frac{i}{2}m\left( \ddot{\theta}_j \sin^2 \theta_j 
 -\dot{\varphi}_j^2\sin{\theta_j}\cos{\theta_j} \right).
\ee
\subee
Due to the regularization term, the higher-order derivatives appear 
in the EOMs and its general solution has more arbitrary constants, 
which naturally enables us to have a solution connecting to both the boundary values $\Omega'$ and $\Omega''$. 
Meanwhile, the terms coming from the regularization introduce 
the imaginary number into the EOMs. 
Hence the corresponding semiclassical path becomes complex in general 
and loses a clear physical interpretation. 
Bloch sphere representation is not applicable to visualize the semiclassical path. 
From a formal correspondence, the Wiener regularization can be regarded 
as a kinetic energy of spins with a pure imaginary mass.

To recover the original action, we take the zero mass limit $m\to 0$. 
For small $m$, the time span $s\in[0,t]$ is divided into 
three characteristic regions~\cite{Alscher:99}: $T_1=[0,m]$, $T_{\rm cl}=[m,t-m]$, and $T_{2}=[t-m,t]$. 
In $T_{\rm cl}$, the mass terms proportional to $m$ become irrelevant and 
the time evolution is essentially driven by the original unregularized EOMs. 
In $T_1$ and $T_2$, the trajectory is strongly hinged by the mass terms 
to match the boundary conditions. 
As a result, in the $m\to 0$ limit, we observe jumps at $s=0$ and $s=t$ 
from the boundary values to the edges of the semiclassical path 
in $T_{\rm cl}\to [0,t]$. 
These jumps give a condition for the values at the boundary $(\bar{\V{\theta}}(0),\bar{\V{\varphi}}(0))$ and 
$(\bar{\V{\theta}}(t),\bar{\V{\varphi}}(t))$, which has a simple explicit form: 
\subbe
 \Leq{matching}
\be
 && \tan\left(\frac{\bar{\theta}_i(0)}{2}\right)e^{i\bar{\varphi}_i(0)}
 =\tan\left(\frac{\theta_i'}{2}\right)e^{i\varphi_i'}, \\ 
 && \tan\left(\frac{\bar{\theta}_i(t)}{2}\right)e^{-i\bar{\varphi}_i(t)}
 =\tan\left(\frac{\theta_i''}{2}\right)e^{-i\varphi_i''}.
\ee
\subee
This condition implies that there can be multiple
semiclassical paths to satisfy \BReq{matching} and that
they can be complex even in the $m\to 0$ limit. 
We note again that, for single-spin systems, 
it was shown in Ref.~\cite{Alscher:99}
that the solution of the unregularized EOMs
\NReq{EOM-theta} under the condition \NReq{matching} gives the exact propagator.

\subsubsection{Solving the boundary value problem}
\Lsec{Solving the boundary}

The boundary value problem becomes well-defined now 
and we can find solutions matching both
the boundary values $\Omega'$ and $\Omega''$ in a generic situation. 
A practical way for solving the problem is to employ 
the following variable transformation~\cite{Alscher:99}: 
\subbe
\be
 &&\zeta_j(s)=\tan\left(\frac{\theta_j(s)}{2}\right)e^{i\varphi_j(s)}, \\
 &&\eta_j(s)=\tan\left(\frac{\theta_j(s)}{2}\right)e^{-i\varphi_j(s)}.
 \Leq{zeta-eta}
\ee
\subee
These variables are, if $(\theta_j(s),\varphi_j(s))$
are real, a stereographic representation of a point on 
the unit sphere projected from the south pole onto the equatorial plane. 
Hence we call them stereographic variables. 
The boundary condition is now written as 
\subbe
 \Leq{boundary-zeta-eta}
\be
 && \zeta_j(0)=\zeta_j' \equiv \tan\left(\frac{\theta_j'}{2}\right)e^{i\varphi_j'}, \\ 
 && \eta_j(t)=\eta_j''\equiv \tan\left(\frac{\theta_j''}{2}\right)e^{-i\varphi_j''},
\ee
\subee
and the remaining boundary values, $\zeta_i(t)$ and $\eta_i(0)$, 
are not specified. 
The spin variables in the Hamiltonian are converted to 
the stereographic variables through the relation
\be
 \langle\theta_j,\varphi_j|\bm{\sigma}_j|\theta_j,\varphi_j\rangle  
 = \frac{1}{1+\zeta_j \eta_j}
 \left(\begin{array}{c}
 \zeta_j+\eta_j \\ -i(\zeta_j-\eta_j) \\ 1-\zeta_j\eta_j
 \end{array}\right),
\ee
and the semiclassical EOMs \NReq{EOM-theta} are
\subbe
 \Leq{EOM-zeta}
\be
 && \dot{\zeta}_j=-i(1+\zeta_j \eta_j)^2\Part{\mathcal{H}}{\eta_j}{}, \\ 
 && \dot{\eta}_j=-i(1+\zeta_j \eta_j)^2\Part{\mathcal{H}}{\zeta_j}{}.
\ee
\subee
Using the solution of the EOMs, $(\bar{\zeta}_j,\bar{\eta}_j)$, 
we can write the  semiclassical action as~\cite{Alscher:99} 
\be
 && e^{S_{\rm cl}[\bar{\V{\zeta}},\bar{\V{\eta}}]}
 = \prod_{j=1}^{N}
 \Blbb
 \sqrt{\frac{(1+\bar{\zeta}_j(0) \bar{\eta}_j(0))(1+\bar{\zeta}_j(t) 
 \bar{\eta}_j(t))}{(1+\zeta_j' \eta_j')(1+\zeta_j'' \eta_j'')}} \no \\ 
 && \times \lb \frac{\zeta_j' \eta_j'\zeta_j'' \eta_j''}{\bar{\zeta}_j(0) 
 \bar{\eta}_j(0)\bar{\zeta}_j(t) \bar{\eta}_j(t)} \rb^{\frac{1}{4}} \Brbb
 \nonumber\\
 && \times 
 \exp\int_0^{t} ds \Blbb\frac{1}{4}\sum_{j=1}^{N}
 \frac{ (1-\bar{\zeta}_j\bar{\eta}_j)(\dot{\bar{\zeta}}_j\bar{\eta}_j
 -\bar{\zeta}_j\dot{\bar{\eta}}_j)}{\bar{\zeta}_j\bar{\eta}_j(1+\bar{\zeta}_j 
 \bar{\eta}_j)} \no \\ 
 && -i\mathcal{H}(\bar{\V{\zeta}},\bar{\V{\eta}},s)\Brbb.
 \Leq{S_cl}
\ee

\subsubsection{Spatially uniform solutions}
\Lsec{Computational difficulty and}

A problem arises when we compute the semiclassical paths satisfying
\BReq{boundary-zeta-eta}.
We need to fix both the initial conditions on $\zeta_i$ 
and the final ones on $\eta_i$.
The initial conditions on $\eta_i$ 
must be selected so as to meet the final conditions. 
This requires us to solve the EOMs many times,
and results in a bottleneck of the present method to
compute the propagator. 
This is because the computational cost for searching such an initial condition
grows exponentially with the number of spins.
Therefore, in practice, we need an assumption that reduces 
the degree of freedom, namely, the computational cost of
searching the initial value of $\eta_i$. 

In the present paper, we assume the spatial uniformity. Our Hamiltonian \NReq{Hamiltonian} has infinite-range interactions and the mean-field ansatz gives the exact result for static systems. Although it is not evident whether the spatial uniformity holds for dynamical systems, we examine this ansatz in the following. The boundary values of $(\zeta_i(s), \eta_i(s))$ are identical for all $i$'s, so that $(\zeta'_i,\eta'_i)=(\zeta',\eta'),$ and $(\zeta''_i,\eta''_i)=(\zeta'',\eta'')$. Then, only two functions, $\zeta(s)$ and $\eta(s)$, are sufficient to describe the dynamics, and the exhaustive search of $\eta(0)$ is now a reasonable task. Moreover, as far as the Loschmidt amplitude is concerned, the initial and final boundary values are common: $\zeta'=\zeta''=\zeta_{\rm b}$ and $\eta'=\eta''=\eta_{\rm b}$. Summarizing these particular conditions, we obtain the explicit formulas of the EOMs as 
\subbe
 \Leq{EOM-uniform}
\be
 && \dot{\zeta}=i\Gamma\lb 1-\zeta^2\rb 
 -2i \zeta \lb h+J\frac{1-\zeta\eta}{1+\zeta\eta}\rb,
 \Leq{EOM-zeta-uniform} \\ 
 && \dot{\eta}=-i\Gamma \lb 1-\eta^2\rb  
 +2i\eta\lb h+J\frac{1-\zeta\eta}{1+\zeta\eta}\rb.~
 \Leq{EOM-eta-uniform}
\ee
\subee
For a given $t$, these EOMs are solved under the conditions 
$\zeta(0)=\zeta_{\rm b}$ and $\eta(t)=\eta_{\rm b}$.
The other boundary values $\zeta(t)$ and $\eta(0)$ are 
not specified and are determined uniquely from the above conditions.
We also note that the relation $\zeta(s)=\eta^*(s)$ does not necessarily 
hold in general.

The solutions $(\bar{\zeta}^{(\nu)}(s),\bar{\eta}^{(\nu)}(s))$ 
are not unique and we can represent the Loschmidt amplitude as 
\be
 \mathcal{L}(t|\Omega_{\rm b})=\bra{\Omega_{\rm b}}e^{-i\hat{\mathcal{H}}t}
 \ket{\Omega_{\rm b}} 
 \sim \sum_{\nu}A_\nu e^{-Nf[\bar{\zeta}^{(\nu)},\bar{\eta}^{(\nu)}]}, \no\\
\ee
where
\be
 && f[\bar{\zeta},\bar{\eta}]
 = -\frac{1}{2}\log\frac{(1+\zeta_{\rm b}\bar{\eta}(0))
 (1+\bar{\zeta}(t)\eta_{\rm b} )}{(1+\zeta_{\rm b} \eta_{\rm b})^2} \no \\ 
 && -i\int_{0}^{t}ds\lb \frac{\Gamma}{2}(\bar{\zeta}
 +\bar{\eta})+h+\frac{J}{2}\frac{1+2\bar{\zeta}\bar{\eta}
 -3\bar{\zeta}^2\bar{\eta}^2}{(1+\bar{\zeta}\bar{\eta})^2}\rb.
 \Leq{f-uniform}
\ee
The time derivative terms are eliminated by performing 
the integration by parts or using the EOMs. 
We also note that the amplitude $A_\nu$ is not important 
to calculate the rate function in Eq.~(\ref{ratef}) at $N\to\infty$.

\subsubsection{Dominant semiclassical paths and a heuristic search procedure}
\Lsec{Dominant semiclassical paths}

Equation \NReq{matching} has a countably infinite number of solutions, and the EOMs do as well. 
Among those many semiclassical solutions, the one that makes the real
part of $f[\bar{\zeta}^{(\nu)},\bar{\eta}^{(\nu)}]$ the smallest gives
the rate function in Eq.~(\ref{ratef}). How can we find such a dominant
solution? The exhaustive search of $\bar{\eta}(0)$ in the whole complex space
is not plausible even under the spatial uniformity. To overcome the
situation, we here give a heuristic procedure  
to obtain such a dominant path. Since the correct initial condition
$\bar{\eta}(0)$ depends on the end time $t$, we hereafter use a notation
$C(t)=\bar{\eta}(0;\bar{\eta}(t)=\eta_{\rm b})$. The basic idea of the heuristic is
starting from a trivial solution at a specific time $t^*$ and extending
it with changing the time $t$ from $t^*$ gradually.  

The first trivial solution is obtained at $t^*=0$, where $C(0)=\eta_{\rm b}$. Then, for a small time step $\Delta t$, $C(\Delta t)$ is obtained as follows. We examine several values as the initial condition for $\bar{\eta}(s)$ around $\eta_{\rm b}$ and solve the EOMs. We select the best one for $C(\Delta t)$ that makes the final value $\bar{\eta}(s=\Delta t)$ closest to $\eta_{\rm b}$. For the next time step $t=2\Delta t$, we examine the values around $C(\Delta t)$ and repeat the same procedures, giving $C(2\Delta t)$. We repeat this procedure until we reach a desired end time $t$, yielding the sequence of the initial condition. We write this sequence as $C_{1}(t)$.

To obtain the second trivial solution, an important observation is that
the dynamics is periodic at most of parameters~\cite{Sciolla:11}. There
exists a specific period $\tau$ and the order parameters at
$t_n=t_0+n\tau$ are identical for $\forall{n} \in \mN$. This implies
that at $t^*=\tau$ the final condition $\bar{\eta}(s=t^*)=\eta_{\rm b}$ is
realized by having the initial condition $\bar{\eta}(0)=\eta_{\rm b}$,
yielding 
$C(t^*=\tau)=\eta_{\rm b}$. Extending $C(t)$ back from $t=\tau$ to $t=0$ based
on the same procedure for $C_1(t)$, we get another sequence of the
initial condition, and write it as $C_{2}(t)$. In \Rfig{C_heuristic}, a
schematic picture of this heuristic is given. 
\begin{figure}[tbp]
\begin{center}
\includegraphics[width=0.99\columnwidth]{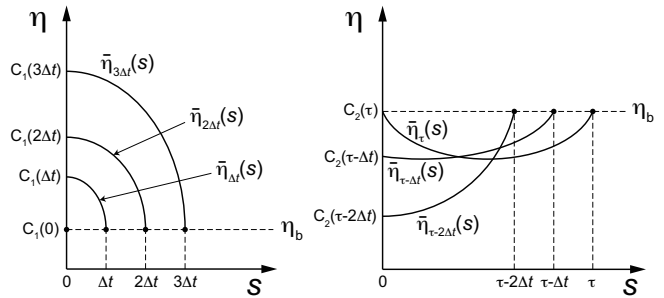}
\caption{Schematic pictures of the heuristic to obtain appropriate
 initial conditions $C_1(t)$ (left panel) and $C_2(t)$ (right panel) of $\bar{\eta}_t$. The complex
 plane of $\eta$ is schematically mapped to the horizontal axis. Here,
 $\bar{\eta}_t(s)$ denotes the semiclassical path satisfying the final
 condition $\bar{\eta}(t)=\eta_b$ for given $t$.
 The initial condition, $\bar{\eta}(0)$ for given $t$ is accordingly
 searched, starting from $t=0$ ($C_1$) or $t=\tau$ ($C_2$).
} 
\Lfig{C_heuristic}
\end{center}
\end{figure}

The question is whether these two sequences of initial conditions,
$C_{1}(t)$ and $C_{2}(t)$, are identical or not. 
If they are different, they give two different semiclassical paths. 
In such a situation, there should be a switch 
between two paths at a certain critical time $t_{\rm c}$ in the period
$[0,\tau]$, that yields a singularity of the Loschmidt amplitude.  
Meanwhile, if they are identical, only one dominant semiclassical path exists 
and is analytic with respect to $t$. 

For longer time $t>\tau$, we repeat the above procedure. 
For the next period $[\tau,2\tau]$, $C_{3}(t)$ is obtained
by extending $C(t)$ from $t=\tau$ to $2\tau$ with the trivial value
$C(\tau)=\eta_{\rm b}$, and $C_{4}(s)$ is given by an extension 
from $t=2\tau$ to $\tau$ with $C(2\tau)=\eta_{\rm b}$.
We note that by construction $C_{2}(s)$ and
$C_{3}(s)$ are continuously connected. 
The solutions for the whole time axis are obtained along this way.

We adopt the above scenario to search the solution.
This may give a wrong result in general, but, 
as far as we have investigated, the result shows a good agreement 
with numerical experiments as we see in the following. 
Our heuristic procedure is constructed under the assumption that 
the system shows a periodic behavior and only one transition at most in one cycle. 
As long as this assumption is true, 
our heuristic can find the correct dominant path. 
For more general cases, e.g. 
spin glasses without periodicity~\cite{Obuchi:12,Takahashi:13}, 
other heuristics should be tailored. 
Investigation of such cases is beyond the scope of this paper and 
will be an interesting future work.

\section{Result}
\label{Result}

We present the results of our semiclassical computation.
We study two cases: 
quenches from  $\Gamma_{\rm i}=\infty$ (\Rsec{Quench from infinity})
and quenches from $\Gamma_{\rm i}=0$ (\Rsec{Quench from zero}). 
The first case is the quench from $\Gamma_{\rm i}=\infty$ to a
finite value $\Gamma_{\rm f}<\infty$, where 
the boundary condition is the ground state at $\Gamma_{\rm i}=\infty$, namely,
$\ket{\Omega_{\rm b}}=\otimes_{i}\ket{\rightarrow}_i$ 
with $\ket{\rightarrow}_i$ being the eigenstate of $\sigma_i^x$ for eigenvalue $+1$. 
The other case is the opposite quench, from $\Gamma_{\rm i}=0$ to $\Gamma_{\rm f}>0$.
We set $h\geq 0+$ and thus $\ket{\Omega_{\rm b}}=\otimes_{i}\ket{\uparrow}_i$. 
Since exact calculation is possible for a quench from 
$\Gamma_{\rm f}=\infty$ to $\Gamma_{\rm i}=0$, we show its result
in \Rsec{Quench from infinity} as well.
We also show the results of numerical studies in \Rsec{Comparison with numerical}
to confirm that the complex semiclassical analysis gives a reasonable result. 

\subsection{Quench from $\Gamma_{\rm i}=\infty$}
\Lsec{Quench from infinity}

In this case, the boundary condition is given by
$(\theta',\varphi')=(\theta'',\varphi'')=(\pi/2,0)$,
that is $(\zeta_{\rm b},\eta_{\rm b})=(1,1)$. 
With this boundary condition, if $h=0$, the state does not evolve and 
the semiclassical path is written as $\bar{\zeta}(s)=\bar{\eta}(s)=1$ for $\forall{s}$.
Hence, we consider the case $h>0$ where, as we show below,
a finite periodicity $0<\tau<\infty$ is present.
In fact, we see several patterns of the rate function and DQPT as well. 
We obtain the corresponding phase diagram.

\subsubsection{A solvable case: $\Gamma_{\rm f}=0$}
\Lsec{A solvable case:}

We first investigate the quench to $\Gamma_{\rm f}=0$. 
In this case, the state is evolved under the classical Ising Hamiltonian 
and an analytical solution of \BReq{EOM-uniform} is available.
We solve the equation under the conditions $\zeta(0)=1$ and $\eta(t)=1$.
Putting the initial condition as $(\zeta(0),\eta(0))=(1,C)$, 
we get the explicit solution of the dynamics as
\subbe
\Leq{solution-G0}
\be
 && \bar{\zeta}(s)=\exp\left(-2is\frac{(1+C)h+(1-C)J}{1+C}\right), \\
 && \bar{\eta}(s)=C\exp\left(2is\frac{(1+C)h+(1-C)J}{1+C}\right). 
\ee
\subee
Then the condition $\eta(t)=1$ gives us 
\be
 C\exp\left(2it\frac{(1+C)h+(1-C)J}{1+C}\right)=1 ,
 \Leq{eta0t-G0}
\ee
which yields $C(t)$.

This example clearly shows the presence of multiple paths satisfying 
the boundary condition. 
As declared in \Rsec{Dominant semiclassical paths},
we investigate two paths associated with the initial conditions $C_1(t)$ and $C_2(t)$, each of which is continuously extended from $C(0)=\eta_{\rm b}=1$ and from $C(\tau)=\eta_{\rm b}=1$ respectively,
where $\tau$ is the period of the dynamics. The period $\tau$ can be obtained by putting $C=1$ and $t=\tau$ in the solution \NReq{solution-G0} as  
\be
 \tau=\frac{\pi}{h}.
\ee
The solutions of \BReq{eta0t-G0} connecting to $C(0)=1$
and $C(\tau)=1$, $C_1(t)$ and $C_2(t)$ respectively, are shown 
in \Rfig{Gf0-h01} for $h/J=0.1$. 
As a reference, the solutions in the next period $[\tau,2\tau]$, 
$C_3$ and $C_4$, are also displayed.  
\begin{figure*}[tbp]
\begin{center}
\includegraphics[width=0.63\columnwidth]{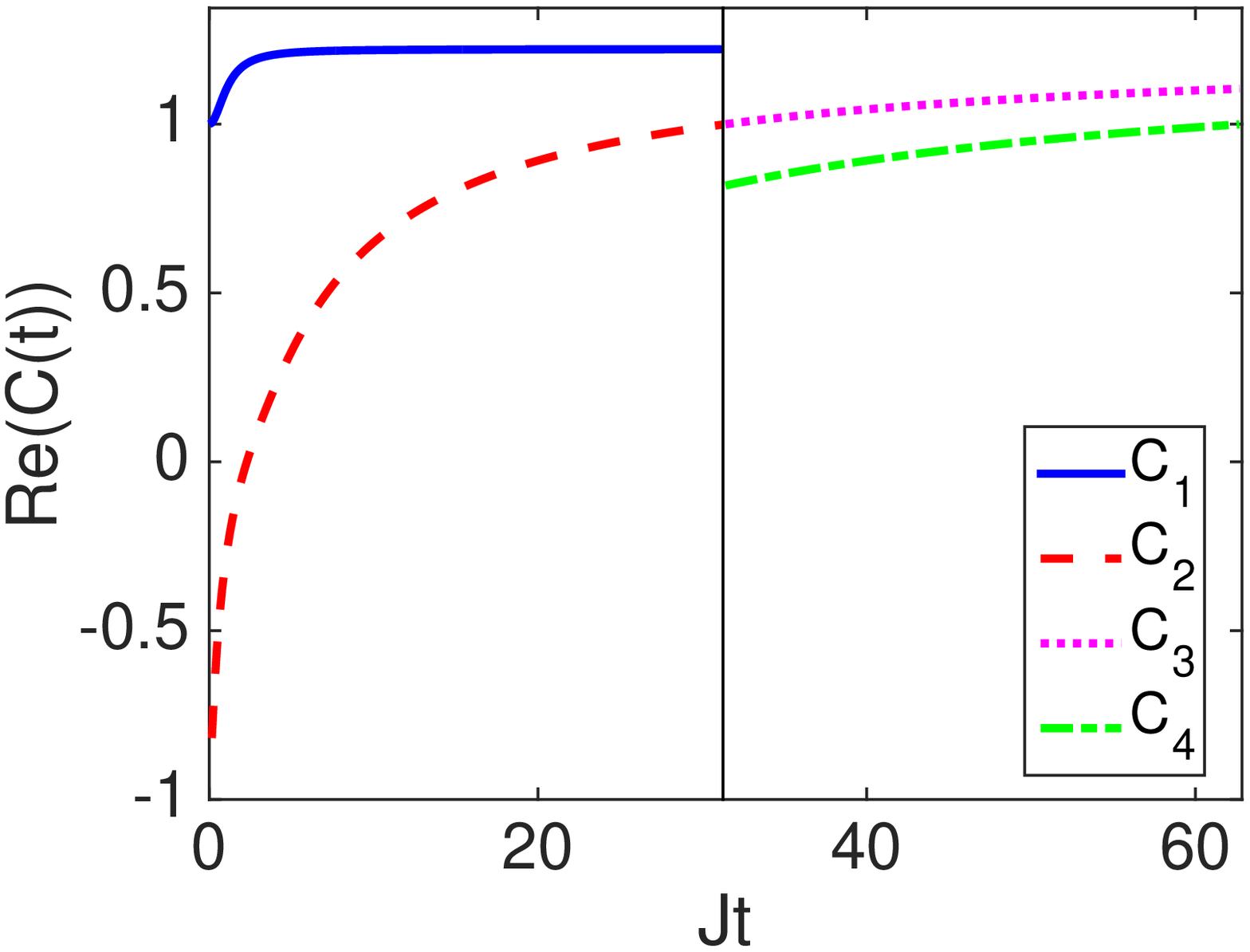}
\includegraphics[width=0.63\columnwidth]{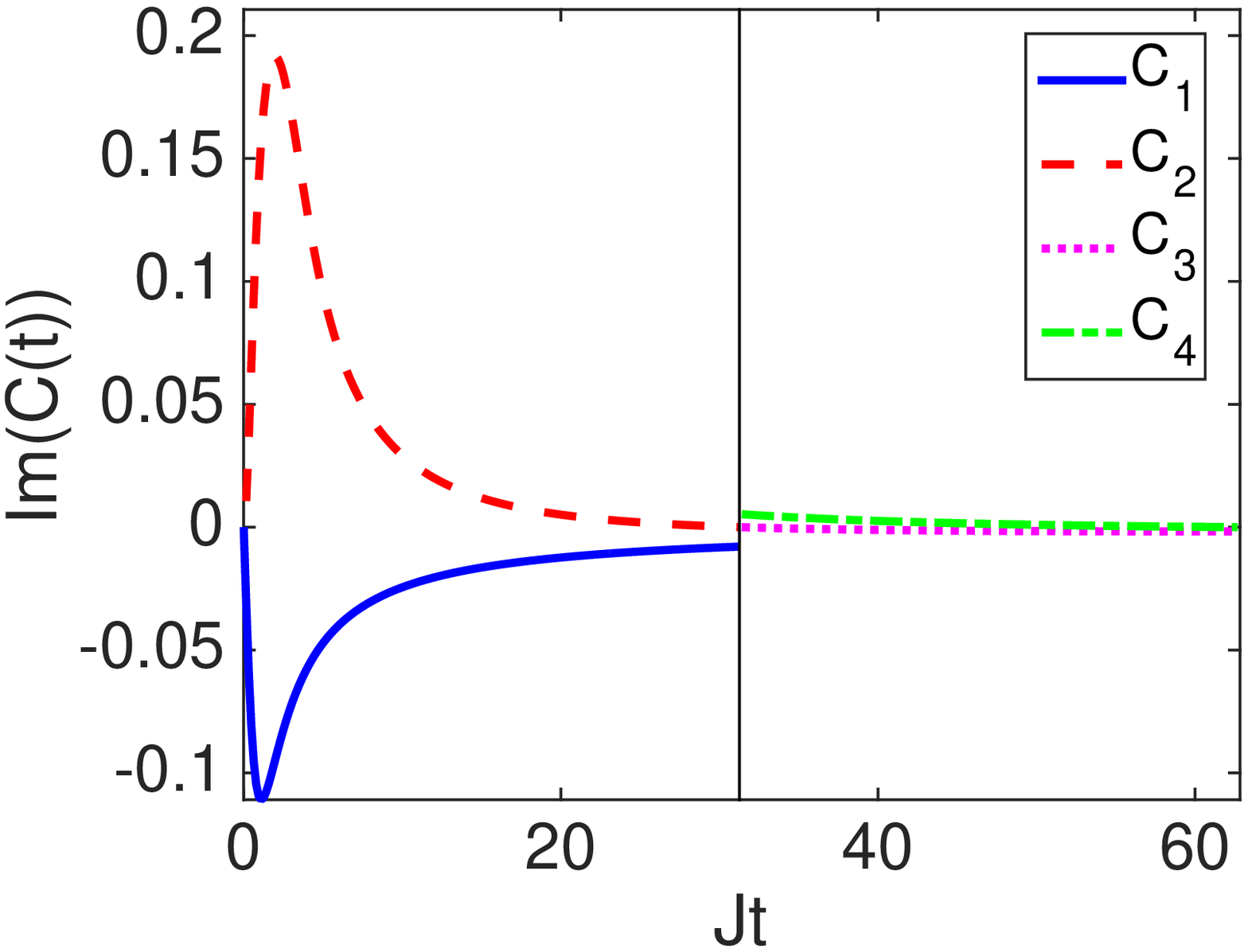}
\includegraphics[width=0.63\columnwidth]{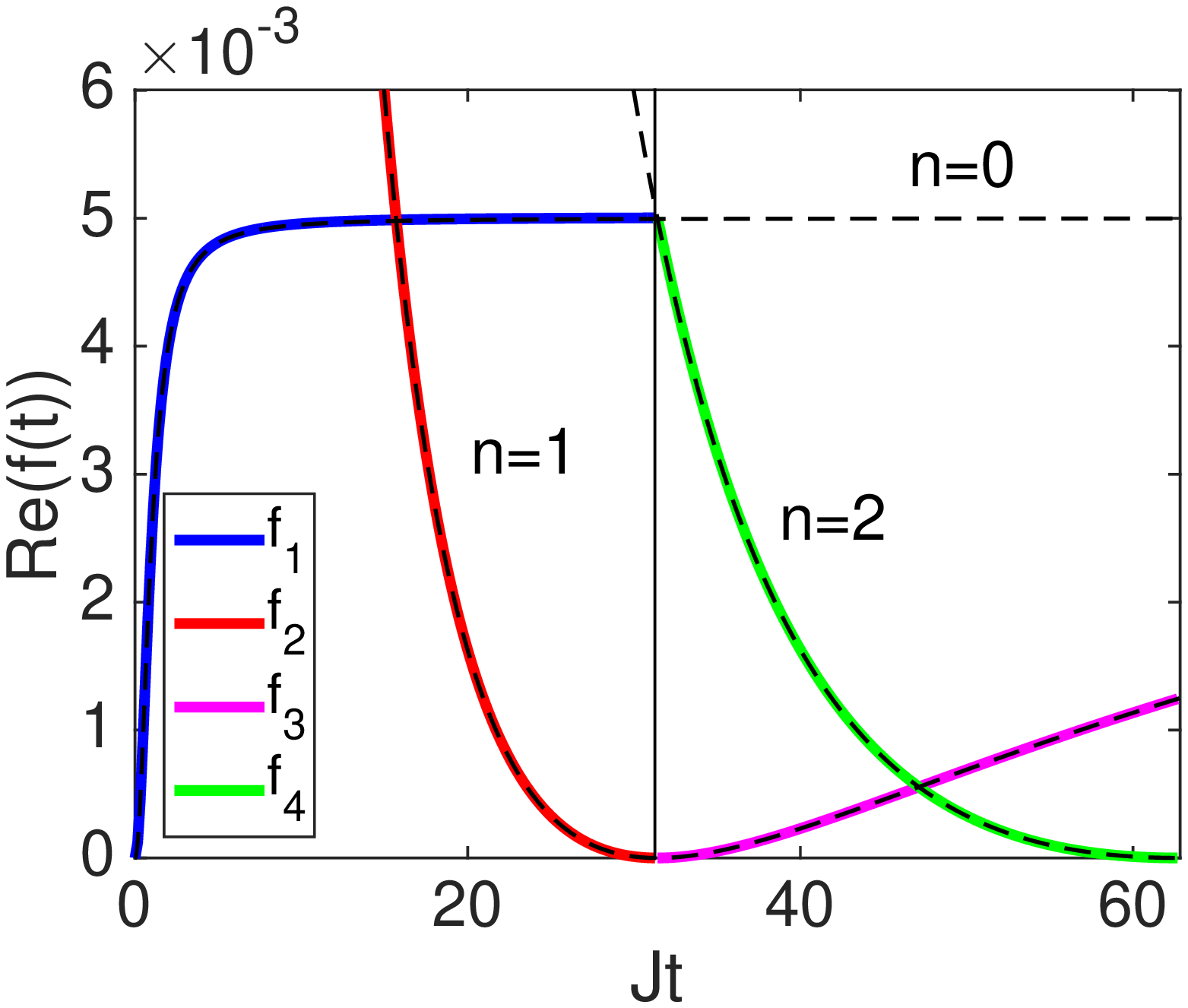}
\caption{(Left, Center) The initial condition $C(t)=\eta(0;\eta(t)=\eta_{\rm b})$ 
satisfying \BReq{matching} 
is plotted against $t$ at $\Gamma_{\rm i}=\infty$, $h/J=0.1$, and $\Gamma_{\rm f}=0$. 
The left and center panels are for the real and imaginary parts, respectively. 
Two periods $[0,\tau]$ and $[\tau,2\tau]$ are focused on, and 
two branches in each period are displayed and the vertical straight line denotes the period $\tau=\pi/h$.  
The period $\tau$ is given by $J\tau=10\pi$ in the present case.
(Right) The real part of the rate function $f_{\rm r}(t)$ 
corresponding to those four branches are shown. 
The analytical solution \NReq{f_r(t;n)} is shown by the dashed lines.
The actual rate function follows the smallest branch at each $t$ 
as in the upper left panel of \Rfig{comp-frominf}.} 
\Lfig{Gf0-h01}
\end{center}
\end{figure*}
Given a time $t$, the rate function $f_k(t)$ corresponding to 
the initial condition $C_k(t)$ is evaluated by inserting \BReq{solution-G0} 
with $C=C_k(t)$ into \BReq{f-uniform} and performing the integration 
with respect to $s$. 
The result is shown in the right panel of \Rfig{Gf0-h01}. 
This exhibits the DQPT at $t=\tau/2$ where the switch from $f_1$ to $f_2$ occurs. 
Similarly, the switch from $f_3$ to $f_4$ occurs at $t=\frac{3}{2}\tau$, 
showing another DQPT.  

Apart from the analytical solution of the EOMs, the rate function itself 
can be computed exactly for the case $\Gamma_{\rm f}=0$. 
Dashed lines in the right panel of \Rfig{Gf0-h01} represent the result. 
In terms of the total spin operator $\hat{S}_{z}=\frac{1}{2}\sum_{i}\sigma_{i}^{z}$, 
our Hamiltonian after the quench is written as
\be
 \hat{\mathcal{H}}=-2\lb \frac{J}{N}\hat{S}_z^2+h\hat{S}_z \rb. 
\ee
The eigenvalue of this Hamiltonian is characterized by that of $\hat{S}_z$ 
denoted by $M$ taking the value $\frac{N}{2}-k$ with $k=0,1,\dots,N$. 
For a given $M$, there are $\binom{N}{k}$ degenerate states. 
Let us define a normalized vector $\ket{\frac{N}{2}-k}$ in this subspace,
which is the equal-weight sum of the $\binom{N}{k}$ basis vectors.
Using this basis, we can write the initial state as
\be
 \ket{\Omega_{\rm b}} = \sum_{k=0}^{N}\lb \frac{1}{2}\rb^{\frac{N}{2}}
 \sqrt{\binom{N}{k}}\ket{\frac{N}{2}-k}.
\ee
Applying the time-evolution operator $e^{-i\hat{\mathcal{H}}t}$
only gives a phase factor for each term. 
The Loschmidt amplitude is written as
\be
 &&\mathcal{L}(t) = \sum_{k=0}^{N}\lb \frac{1}{2}\rb^{N}\binom{N}{k} \nonumber\\
 &&\times \exp \left\{
 {2i\left[ \frac{J}{N}\lb \frac{N}{2}-k\rb^2+h\lb \frac{N}{2}-k\rb \right] t}
 \right\}.
\ee
Using an approximation valid for $N\gg 1$
\be
 \lb \frac{1}{2}\rb^{N}\binom{N}{k}
 \sim \sqrt{\frac{2}{\pi N}}\exp\left[-\frac{2}{N}\lb k-\frac{N}{2}\rb^2\right],
\ee
we write the amplitude as 
\be
 \mathcal{L}(t) &\sim& 
 \sqrt{\frac{2}{\pi N}}\sum_{k=-\frac{N}{2},-\frac{N}{2}+1,\dots,\frac{N}{2}}
 \exp\biggl[-\frac{2}{N}(1-iJt)k^2 \nonumber\\
 && 
 -2ihtk\biggr].
\ee
For $N\gg 1$ the range of the sum can be safely extended from
$k=-\infty$ to $\infty$ to yield 
\be
 \mathcal{L}(t) &\sim& \sqrt{\frac{2}{\pi N}}\sum_{n=-\infty}^{\infty}
 \int_{-\infty}^{\infty} d\phi \nonumber\\ 
 &&\times \exp\left[-\frac{2}{N}(1-iJt)\phi^2-2iht\phi +2i\pi n \phi \right] \no \\ 
 &=& \sqrt{\frac{1}{1-iJt}}\sum_{n=-\infty}^{\infty}
 \exp\left[-N\frac{(ht-\pi n)^2}{2(1-iJt)}\right],
 \Leq{L-Gamma0}
\ee
where the Poisson summation formula is used in the first line~\cite{NishimoriBook}.
We then define $n$ that minimizes $f_{\rm r}(t;n)$ 
as $n^*(t)=\argmin_{n\in\mZ}f_{\rm r}(t;n)$, where
\be
 f_{\rm r}(t;n)=\frac{(ht-\pi n)^2}{2(1+J^2t^2)}.
 \Leq{f_r(t;n)}
\ee
The contribution from $n=n^*$ dominates the sum in \BReq{L-Gamma0} and 
we obtain the real part of the rate function as 
$f_{\rm r}(t) = f_{\rm r}(t,n^*(t))$.
$f_{\rm r}(t)$ exhibits singularities 
because $n^*(t)$ changes discretely as $t$ grows. 
Thus the transition time $t_{\rm c}$ is obtained by equating two neighboring 
values $f_{\rm r}(t;n)=f_{\rm r}(t;n+1)$ as
\be
 t_{\rm c}(n)=\frac{\pi}{h}\lb n+\frac{1}{2}\rb.
\Leq{t_c}
\ee
Hence the period is given by $\tau=\pi/h$. The branches of $f_{\rm r}(t;n)$ with $n=0,1,2$ are shown in the right panel of \Rfig{Gf0-h01}, which exhibits the perfect agreement with $f_k(t)$ evaluated by the integration in \BReq{f-uniform} with the solution of EOMs \NReq{solution-G0} and $C_k(t)$.

Two noteworthy consequences are provided by this analytical solution. One is that the DQPT always exists for any $h>0$, while it does not for $h=0$. Some earlier works have pointed out that a DQPT appears when quench crosses an equilibrium quantum phase transition~\cite{Heyl:13,Karrasch:13}. The present results reveal the existence of the opposite situation. The other is that the real part of
the rate function $f_{\rm r}(t)=f_{\rm r}(t;n^{*}(t))$ shrinks by the speed of $O(t^{-2})$ as $t$ grows and finally vanishes in the limit $t\to \infty$. The vanishing rate function may be thought to imply
$|\mathcal{L}(t)|\to 1$, but this is not the case because of the presence of the factor $1/\sqrt{1-iJt}$ in \BReq{L-Gamma0}. The modulus of this factor decreases as $t$ grows, so that $|\mathcal{L}(t)|$ goes to zero. This implies that there exists a crossover time $t_{\times}$ determined by comparing the $O(1)$ factor and the exponentially scaling one $e^{-Nf}$. For $t>t_{\times}$ the $O(1)$ factor dominates the Loschmidt amplitude. However, the crossover time $t_{\times}$ is expected to a unbounded increasing function of $N$. Hence in the large size limit our computation of the rate function is meaningful in the whole time region.  

\subsubsection{General $\Gamma_{\rm f}>0$}
\Lsec{General}

Let us proceed to general final values $\Gamma_{\rm f}>0$. 
The analytical solution of the EOMs is not available in this case.
Hence we numerically search the initial conditions $C_1(t)$ and $C_2(t)$, 
and evaluate the corresponding rate functions $f_1(t)$ and $f_2(t)$.  

Our heuristic procedure starts from evaluating 
the period $\tau$ of the dynamics. 
For this purpose, we run the numerical simulation of the EOMs \NReq{EOM-uniform} 
using the naive initial condition, 
$\zeta(0)=\zeta_{\rm b}=1$ and $\eta(0)=\eta_{\rm b}=1$. 
We employ the Runge-Kutta method of the fourth order. 
As an example, we show the result for the case 
with $h/J=0.1$ and $\Gamma_{\rm f}/J=0.6$ in the left panel of \Rfig{Gf06-h01}.
\begin{figure*}[tbp]
\begin{center}
\includegraphics[width=0.67\columnwidth]{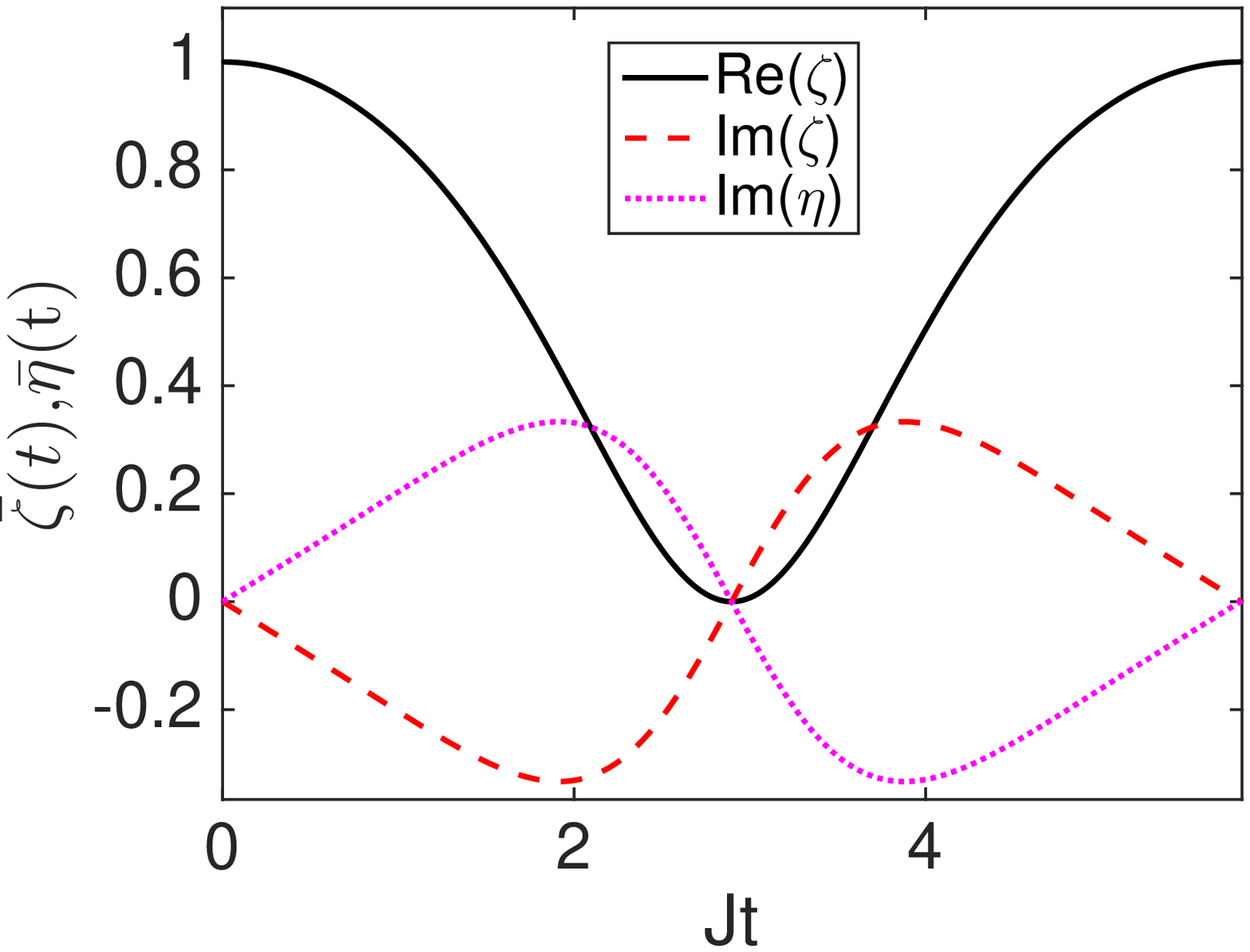}
\includegraphics[width=0.67\columnwidth]{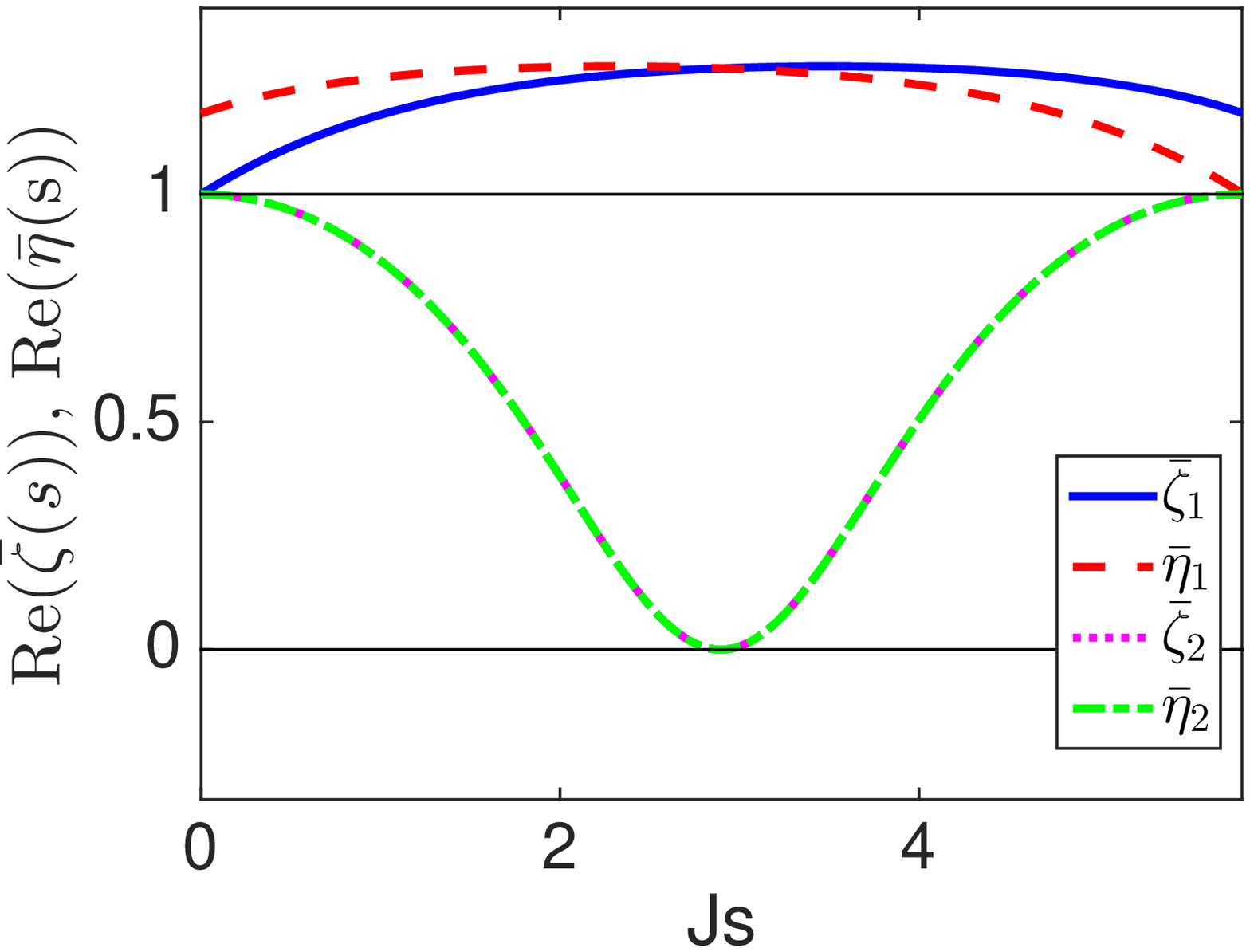}
\includegraphics[width=0.67\columnwidth]{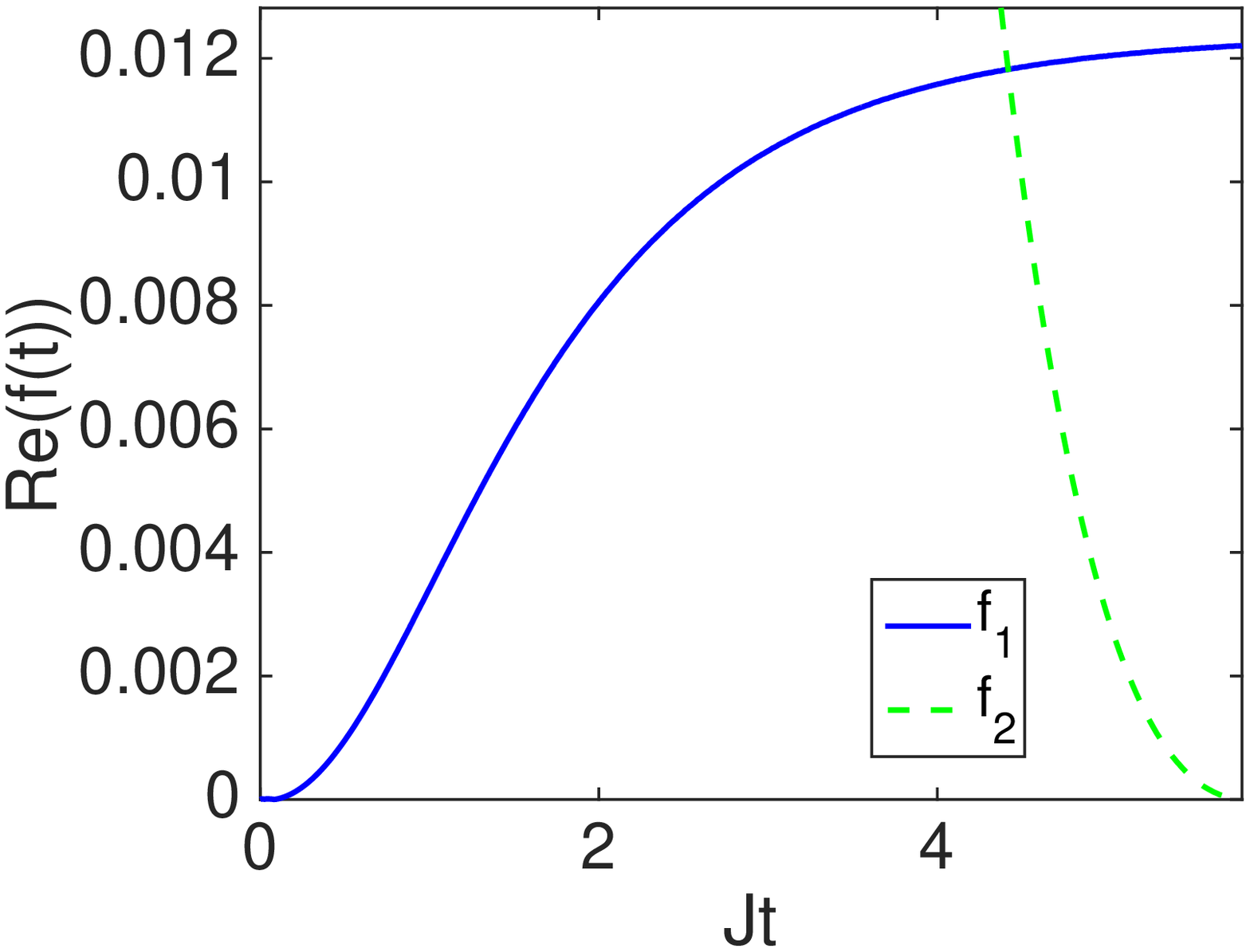}
\caption{Semiclassical paths and the rate functions at 
$\Gamma_{\rm i}=\infty$, $h/J=0.1$, and $\Gamma_{\rm f}/J=0.6$. 
(Left) The paths with the initial condition 
$\zeta(0)=\zeta_{\rm b}$ and $\eta(0)=\eta_{\rm b}$. 
The period $J\tau\approx 5.8$ can be read off. 
The real part of $\eta$ is omitted because $\Re\{\eta(s)\}=\Re\{\zeta(s)\}$.
(Center) Given the end time $t=\tau$, the real parts of semiclassical paths 
with the modified initial conditions $\eta(0)=C(t)$ to satisfy \BReq{matching} are plotted against the dummy time $s$. Two different paths corresponding to different initial conditions, $C_1$ and $C_2$, are shown. The real parts of $\zeta_2$ and $\eta_2$ are identical and are overlapping. 
As a guide to the eye, two horizontal straight lines are drawn at unity and zero.
(Right) Two branches of the rate function $f_1(t)$ and $f_2(t)$. 
A DQPT occurs around $Jt_{\rm c}\approx 4.42$.}
\Lfig{Gf06-h01}
\end{center}
\end{figure*}
The period $J\tau\approx 5.8$ is easily read from this panel. 
We again stress that this dynamics with the naive condition $\eta(0)=\eta_{\rm b}=1$ 
does not satisfy the boundary condition
\NReq{matching} for a generic end time $t$. 
Given an end time $t$, we need to estimate the appropriate initial condition 
$\eta(0)=C(t)$, and then compute the path $(\bar{\zeta}(s),\bar{\eta}(s))$. 
As a result, the semiclassical paths satisfying \BReq{matching} are 
very different from the naive ones. 
Putting the end time as $Jt=J\tau \approx 5.8$, we plot the real parts of such paths in
the center panel of the same figure. 
As explained in \Rsec{Dominant semiclassical paths}, we have 
two different sequences of the initial conditions, yielding two different paths
$(\bar{\zeta}_1(s),\bar{\eta}_1(s))$ and $(\bar{\zeta}_2(s),\bar{\eta}_2(s))$. 
Both paths satisfy $\bar{\zeta}(0)=\zeta_{\rm b}=1$ and $\bar{\eta(t)}=\eta_{\rm b}=1$ as they should. 
The real parts of the corresponding two rate functions 
are plotted in the right panel. 
The smaller branch at each time corresponds to the true rate function, 
leading to the DQPT at $Jt_{\rm c}\approx 4.42$ as a switch from $f_1$ to $f_2$. 
Note that this panel is plotted against the end time $t$ 
while the center one is plotted against the dummy time $s$, 
given the end time $t=\tau$.  

The DQPT observed here has the nature of the first order transition, 
in a sense that the first order time derivative of the rate function 
jumps at the transition time. 
By examining the several parameters, we have realized
that this first order nature tends to be stronger as $\Gamma_{\rm f}$ increases,
but suddenly vanishes at a certain critical value $\Gamma_{\rm fc}(h)$. 
For $\Gamma>\Gamma_{\rm fc}(h)$, the curve of the rate function has 
a smooth peak without singularity. 
In \Rfig{Gf15_16-h01}, we plot $C_1,C_2,f_1,$ and $f_2$ for 
$h/J=0.1$ with slightly different two values of 
$\Gamma_{\rm f}$, $\Gamma_{\rm f}/J=1.5$ and $1.6$. 
\begin{figure}[tbp]
\begin{center}
\includegraphics[width=0.47\columnwidth]{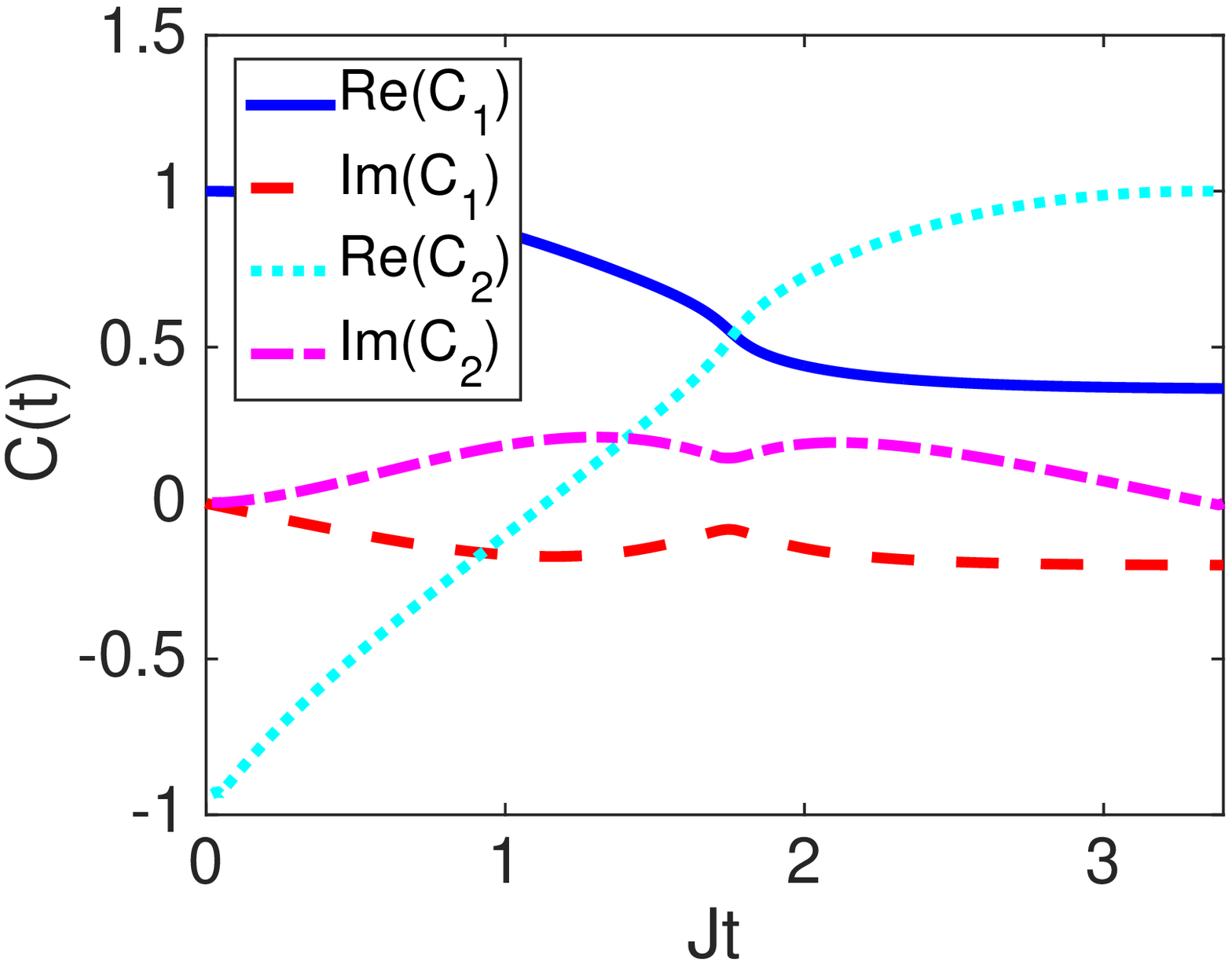}
\includegraphics[width=0.47\columnwidth]{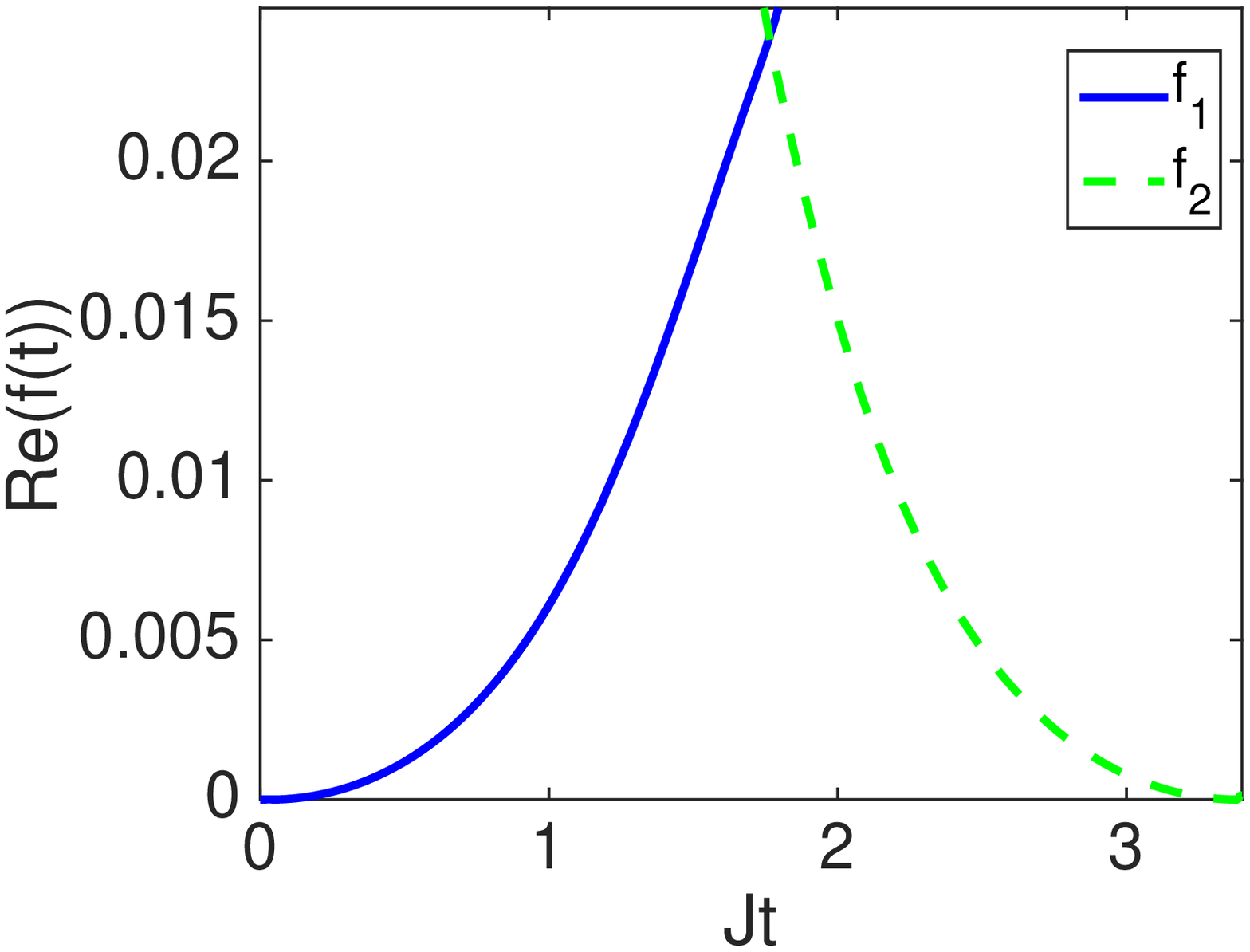}
\includegraphics[width=0.47\columnwidth]{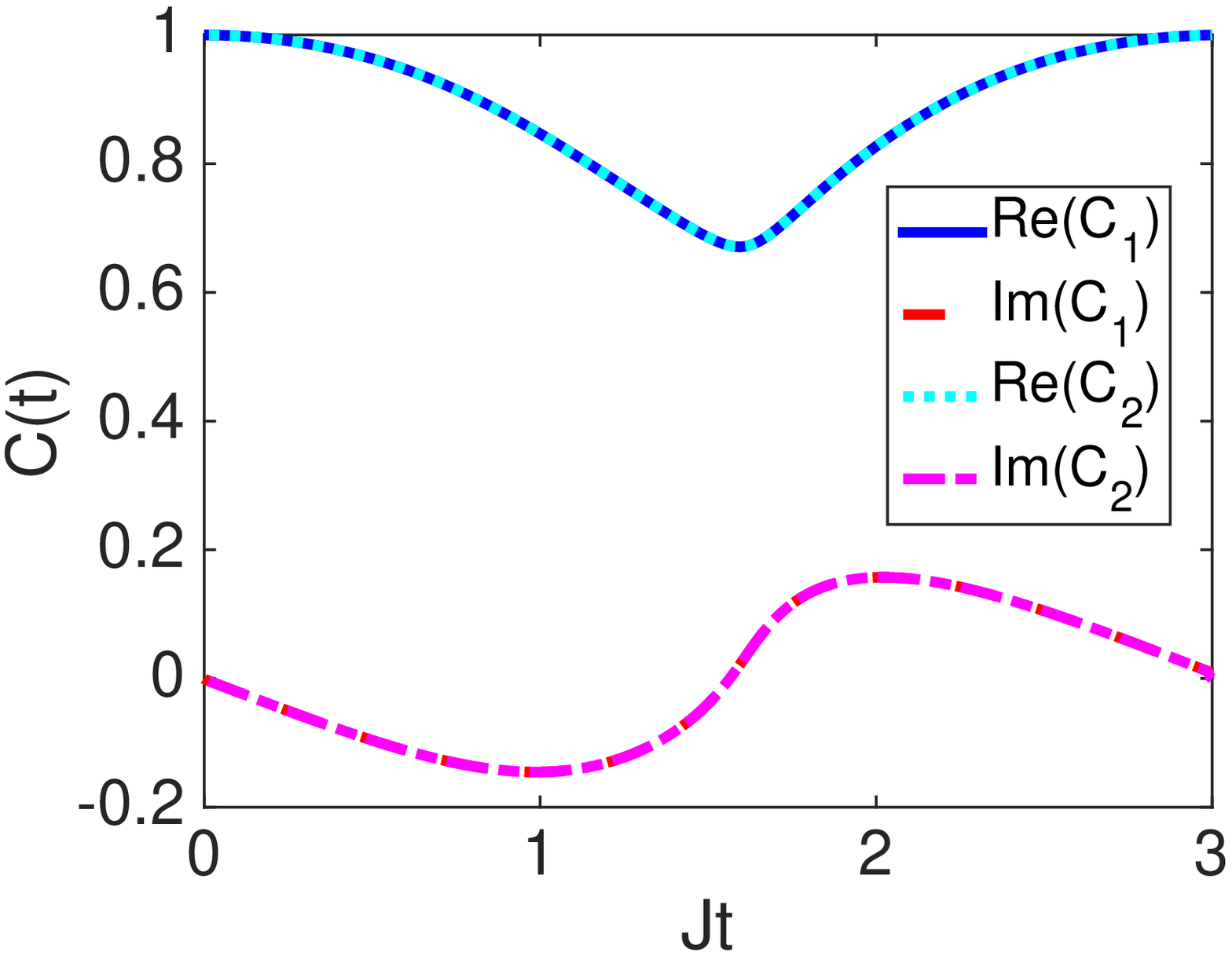}
\includegraphics[width=0.47\columnwidth]{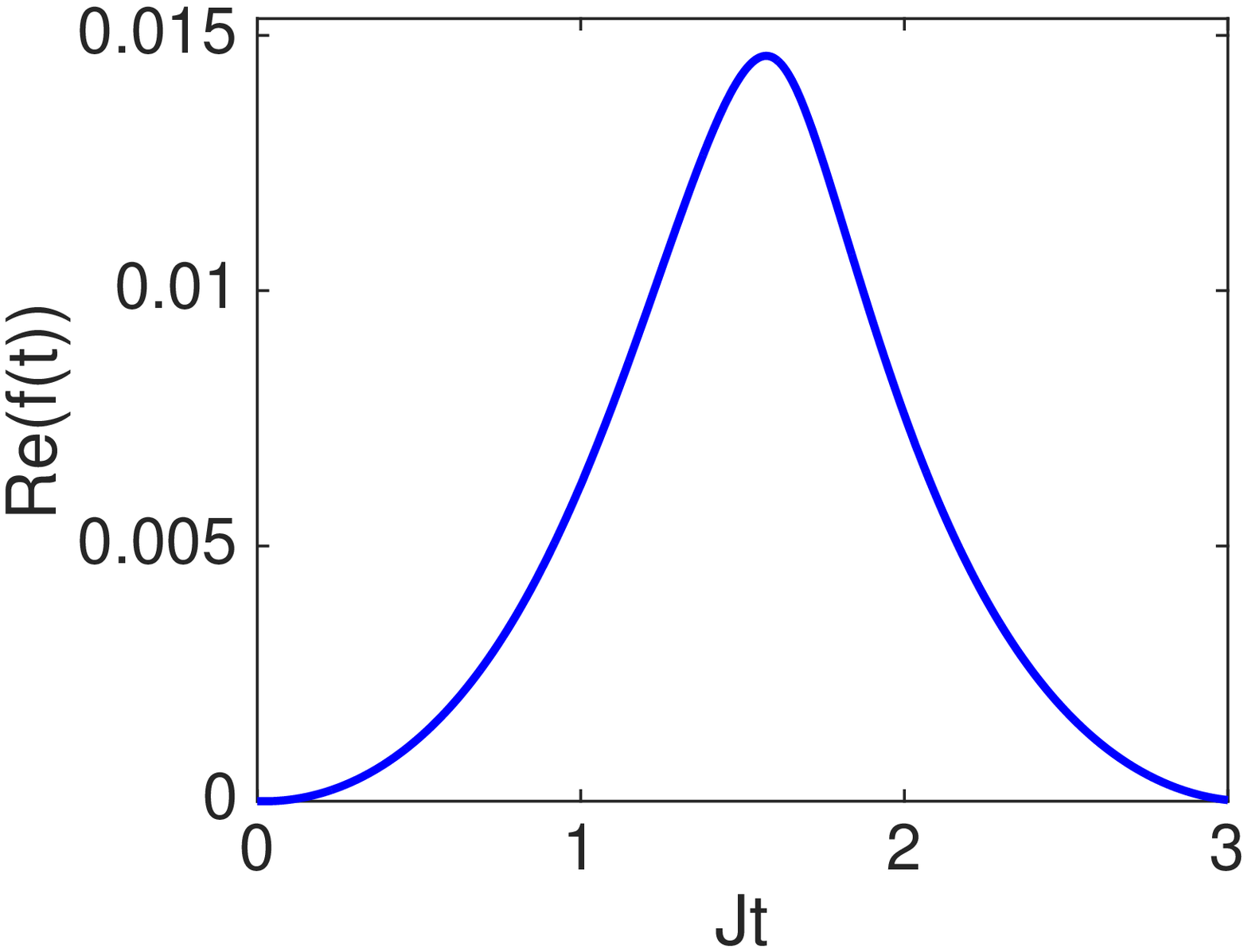}
\caption{Plot of the initial conditions $C_1$ and $C_2$ (Left) and the corresponding rate functions $f_1$ and $f_2$ (Right) at $\Gamma_{\rm i}=\infty$ and $h/J=0.1$ with $\Gamma_{\rm f}/J=1.5$ (Top) and $\Gamma_{\rm f}/J=1.6$ (Bottom). For $\Gamma_{\rm f}/J=1.5$, two different branches exist and a DQPT occurs at $Jt_{\rm c}\approx 1.76$, while they are merged and only one analytic curve is present for $\Gamma_{\rm f}/J=1.6$. }
\Lfig{Gf15_16-h01}
\end{center}
\end{figure}
They clearly show that the critical value $\Gamma_{\rm fc}(h)$ is present
between these two values of $\Gamma_{\rm f}$. 
In the same way, computing the rate function in a range of $h$ and $\Gamma_{\rm f}$, 
we draw a phase diagram in the case of quench 
from $\Gamma_{\rm i}=\infty$ in \Rfig{PD_XP}.
\begin{figure}[tbp]
\begin{center}
\includegraphics[width=0.8\columnwidth]{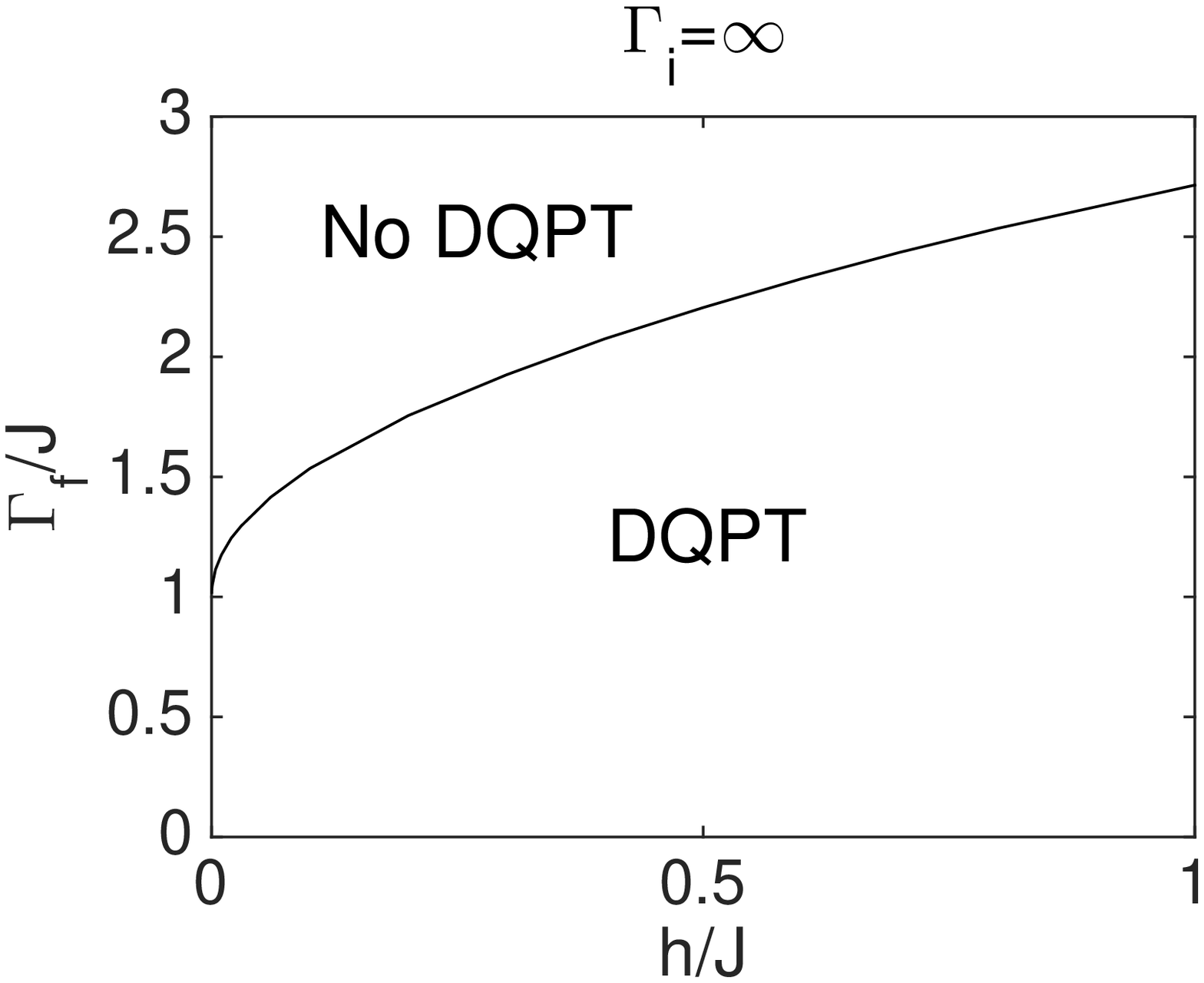}
\caption{The phase diagram for the quench from $\Gamma_{\rm i}=\infty$. The phase boundary line is the interpolation of the data points}
\Lfig{PD_XP}
\end{center}
\end{figure}
The phase boundary approaches to the equilibrium transition point 
$\Gamma_{\rm c}=J$ in the limit $h\to 0$. 
This is reasonable because the period of the dynamics $\tau$ diverges 
as $h \to 0$ at $\Gamma_{\rm f}<\Gamma_{\rm c}$, 
and DQPTs do not exist according to the present scenario.

\subsection{Quench from $\Gamma_{\rm i}=0$}
\Lsec{Quench from zero}

We next study the opposite quench from $\Gamma_{\rm i}=0$. The boundary condition is now given by $\zeta_{\rm b}=\eta_{\rm b}=0$. 

As in the previous case, the numerical search of $C_1(t)$ and $C_2(t)$
brings the behavior of the rate function and the DQPT in this
case. However the results are rather different.
In the previous case, the DQPT was the first order like and there was a
prominent cusp in a period $[0,\tau]$. When going across the DQPT
boundary, the cusp turned into a smooth peak and the bifurcation or merge
of the two initial conditions $C_1(t)$ and $C_2(t)$ occurs in the middle
of the period $[0,\tau]$. For the quench from $\Gamma_{\rm i}=0$,
however, this is not the case and the DQPT emerges in a more delicate
form.  

Figure \NRfig{Gf06_07-h01} is the plots of $C_1,C_2,f_1,$ and $f_2$ for $h/J=0.1$ with slightly different two values of $\Gamma_{\rm f}$, $\Gamma_{\rm f}/J=0.6$ and $0.7$. 
\begin{figure}[tbp]
\begin{center}
\includegraphics[width=0.45\columnwidth]{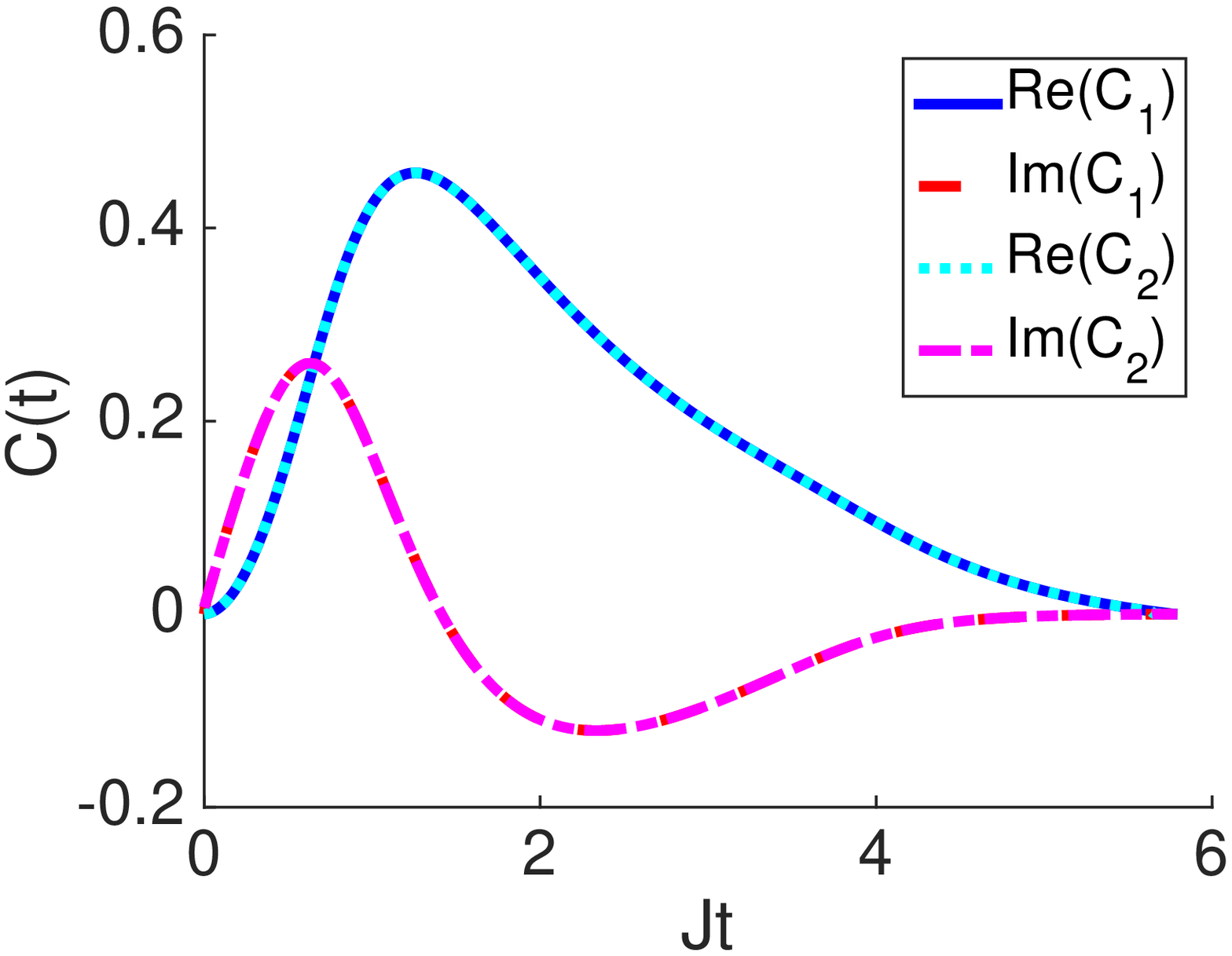}
\includegraphics[width=0.45\columnwidth]{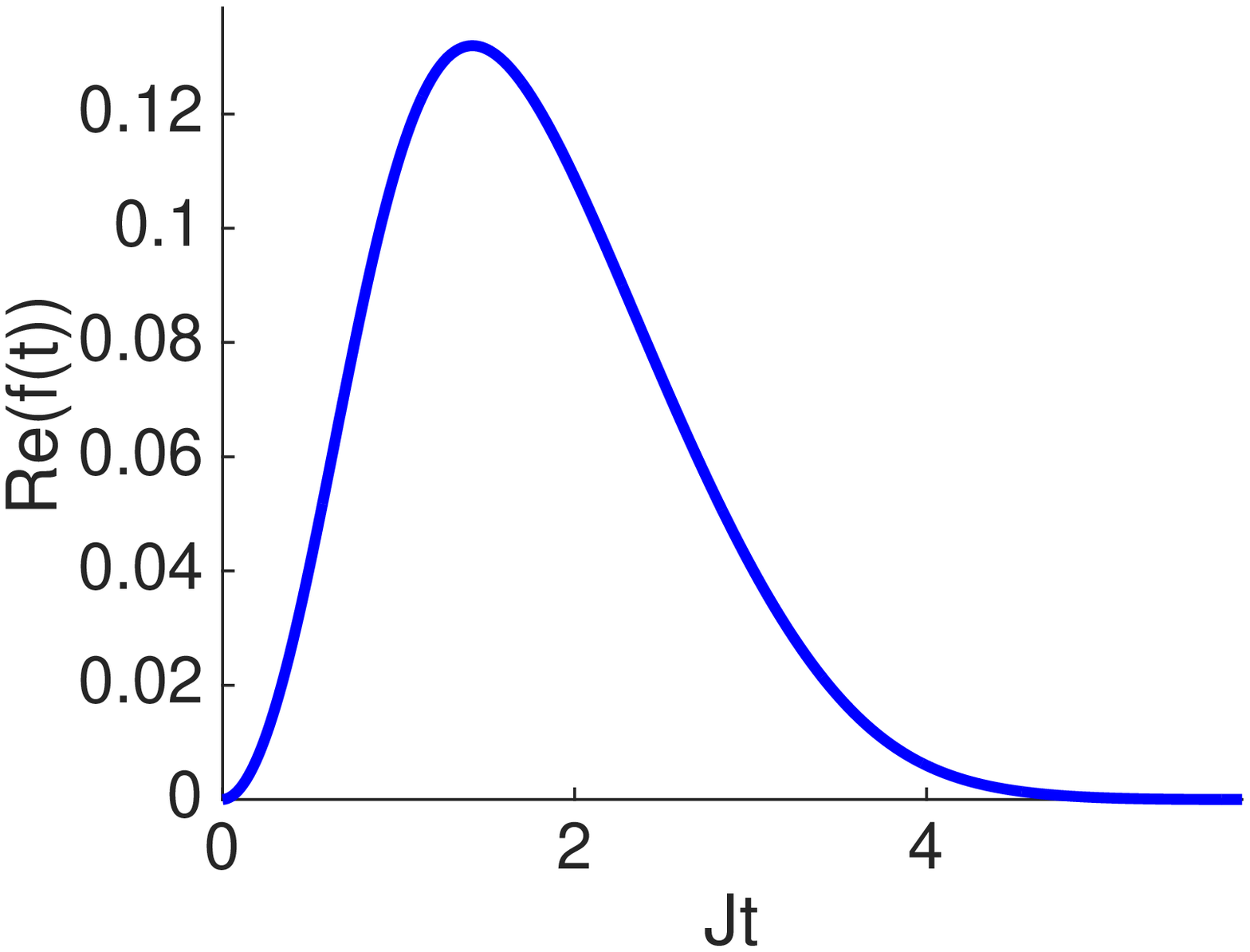}
\includegraphics[width=0.45\columnwidth]{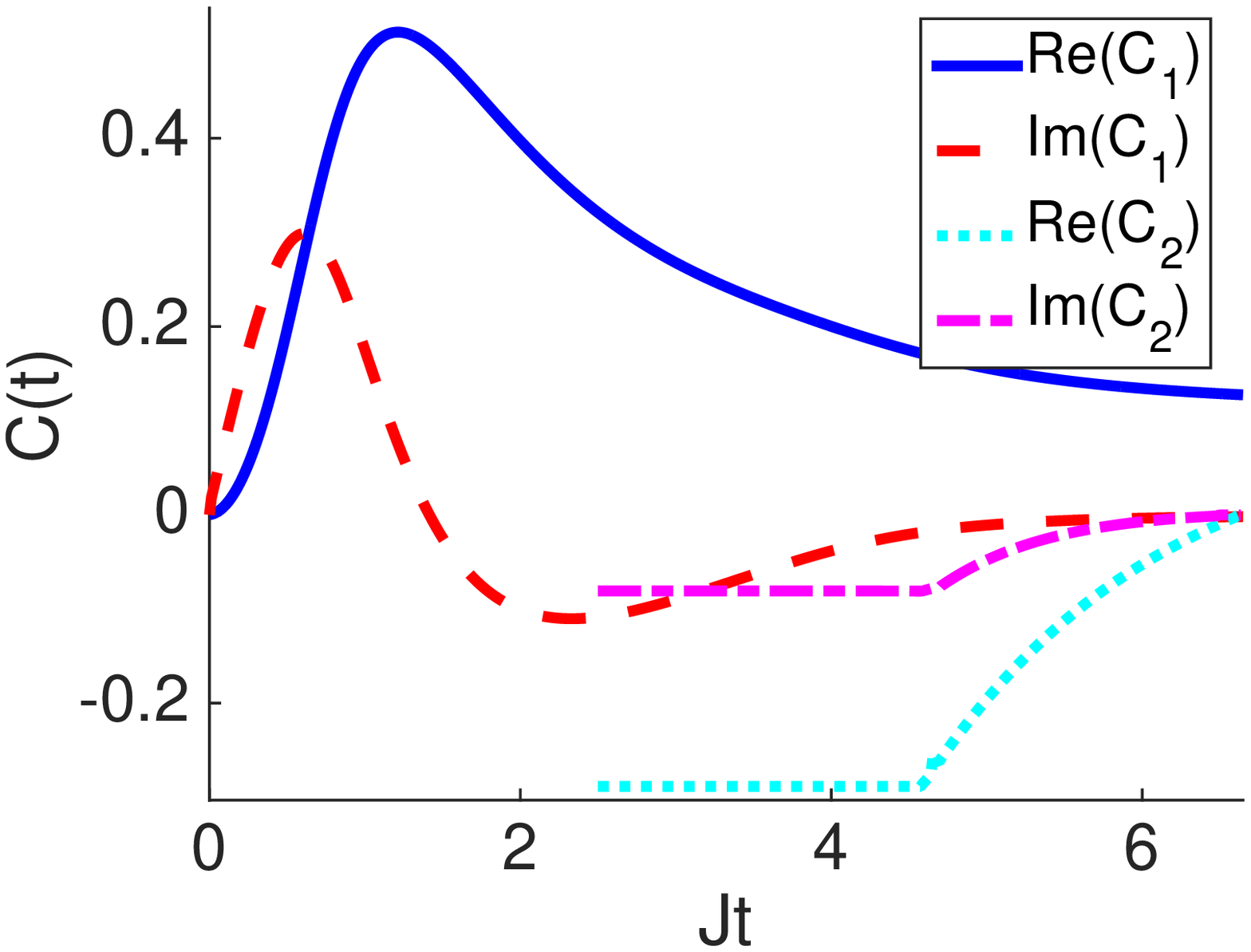}
\includegraphics[width=0.45\columnwidth]{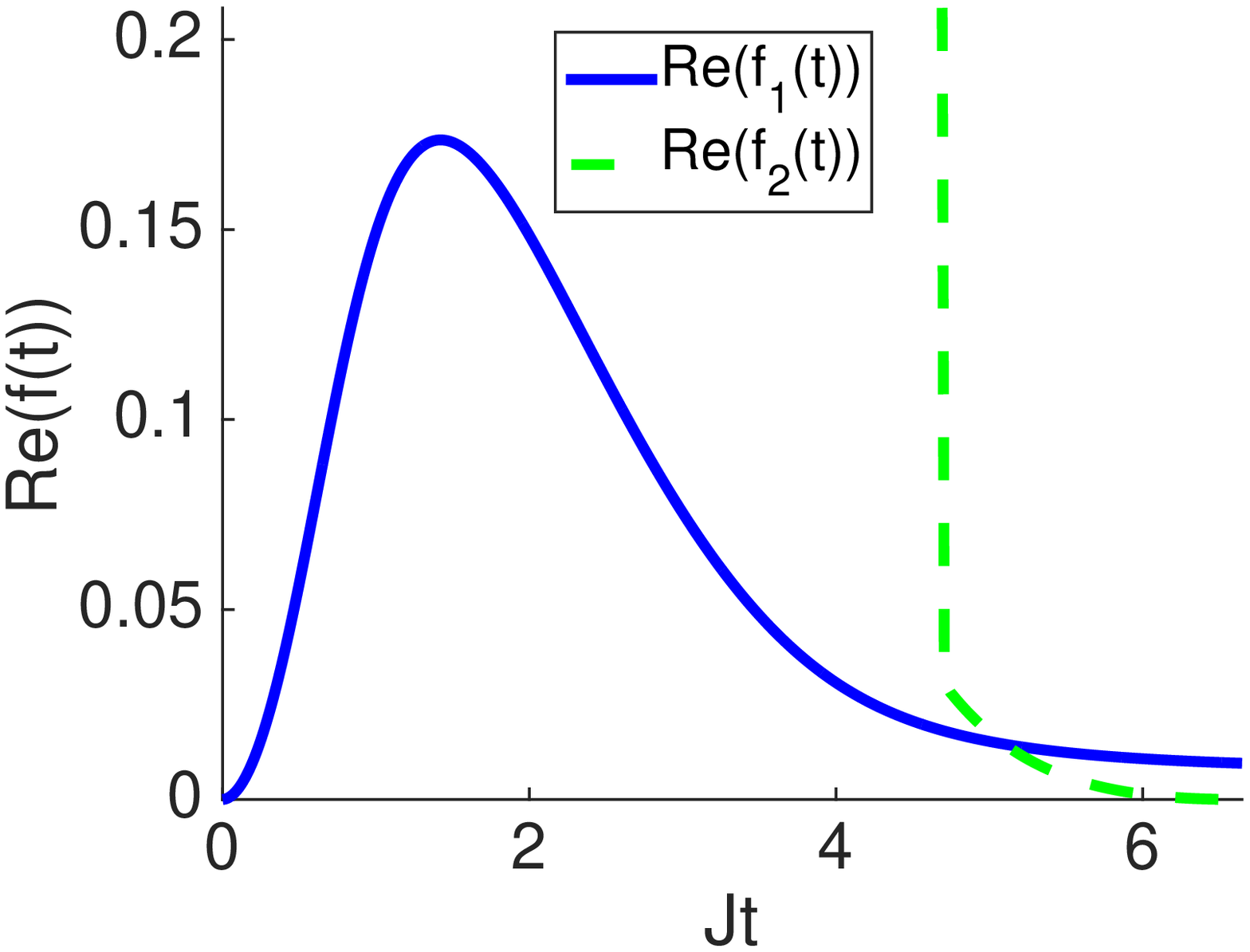}
\caption{Plot of the initial conditions $C_1$ and $C_2$ (Left) and the corresponding rate functions $f_1$ and $f_2$ (Right) at $\Gamma_{\rm i}=0$ and $h/J=0.1$ with $\Gamma_{\rm f}/J=0.6$ (Top) and $\Gamma_{\rm f}/J=0.7$ (Bottom). The period $\tau$ of the dynamics is $J\tau \approx 5.8$ and $J\tau \approx 6.6$ for $\Gamma_{\rm f}/J=0.6$ and $\Gamma_{\rm f}/J=0.7$, respectively. Comparing $\Gamma_{\rm f}/J=0.6$ and $0.7$, we can see a new branch emerges around $t\approx \tau$, which gives a DQPT at a very close time to $\tau$. For the bottom right panel, $f_{\rm r}$ coming from the plateau region of $C_2(t)$ is out of the range in the shown scale, meaning that it is irrelevant for the DQPT.} 
\Lfig{Gf06_07-h01}
\end{center}
\end{figure}
This figure demonstrates that the bifurcation of the two initial conditions $C_1(t)$ and $C_2(t)$ occurs around $t\approx \tau$ in a rather continuous manner. As a result, discriminating the two branches of the solution is harder than the quench from $\Gamma_{\rm i}=\infty$. This tendency holds for the range of $h$ and $\Gamma_{\rm f}$ we have searched, which requires us to conduct a more precise numerics to obtain the phase diagram. Moreover, as we see from the bottom panels ($\Gamma_{\rm f}/J=0.7$), $C_2(t)$ tends to show a rather singular behavior: a smooth curve suddenly changes into a plateau as $t$ decreases and finally it vanishes for small $t$. Although we cannot completely reject a possibility that these behaviors are caused by certain numerical errors, we have carefully checked and confirmed that the shown $C_2(t)$ satisfies the boundary condition in a good precision for the intermediate and large $t$ region, and no branches exists continuously connected to the plateau for small $t$. Hence, these singular behaviors are expected to be true. Fortunately, they are irrelevant for locating the DQPT point since the DQPT occurs at larger $t$ where no pathological behavior appears. The resultant phase diagram for $\Gamma_{\rm i}=0$ is given in \Rfig{PD_ZP}.
\begin{figure}[tbp]
\begin{center}
\includegraphics[width=0.8\columnwidth]{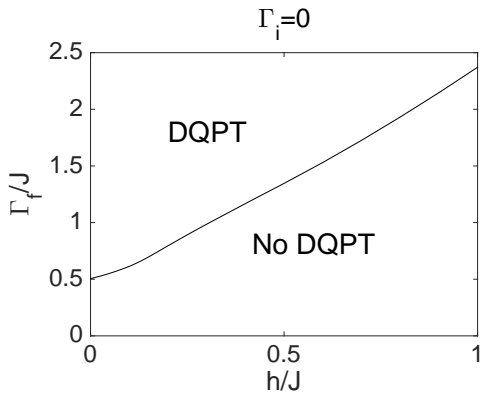}
\caption{The phase diagram for the quench from $\Gamma_{\rm i}=0$. }
\Lfig{PD_ZP}
\end{center}
\end{figure}
The phase boundary $\Gamma_{\rm fc}(h)$ approaches to $\Gamma_{\rm d}/J=1/2$ in
the limit $h\to 0+$. This is reasonable because $\Gamma_{\rm d}$ is the dynamical transition point of an order parameter, $m_z=\bra{\Psi(t)}\sigma^z \ket{\Psi(t)} $ where $\ket{\Psi(t)}=e^{-it\hat{\mathcal{H}}}\ket{\Omega_{\rm b}}$, with this particular choice of $\Gamma_{\rm i}$~\cite{Sciolla:11}. Upon approaching to $\Gamma_{\rm d}$, the period $\tau$ of the dynamics 
diverges and DQPTs should vanish. Note that this dynamical value $\Gamma_{\rm d}$ does not have any meaning for the equilibrium transition. This is in contrast to the quench from $\Gamma_{\rm i}=\infty$ where the equilibrium transition point $\Gamma_{\rm c}$ works as the DQPT transition point at $h=0$. 

Unlike in the $\Gamma_{\rm i}=\infty$ case, the dynamics does not stop even at $h=0$. 
This enables us to see an interesting behavior of the Loschmidt amplitude at $h=0$ and $\Gamma_{\rm f}=\Gamma_{\rm d}$. This point is on the separatrix in the phase space and the order parameter monotonically decreases as $t$ grows. No periodicity exists (or $\tau=\infty$). Hence, we only examine the first sequence of the initial condition for $\eta(s)$ and $C_1(t)$, and compute the corresponding rate function. The result is shown in \Rfig{separatrix}. 
\begin{figure*}[tbp]
\begin{center}
\includegraphics[width=0.63\columnwidth]{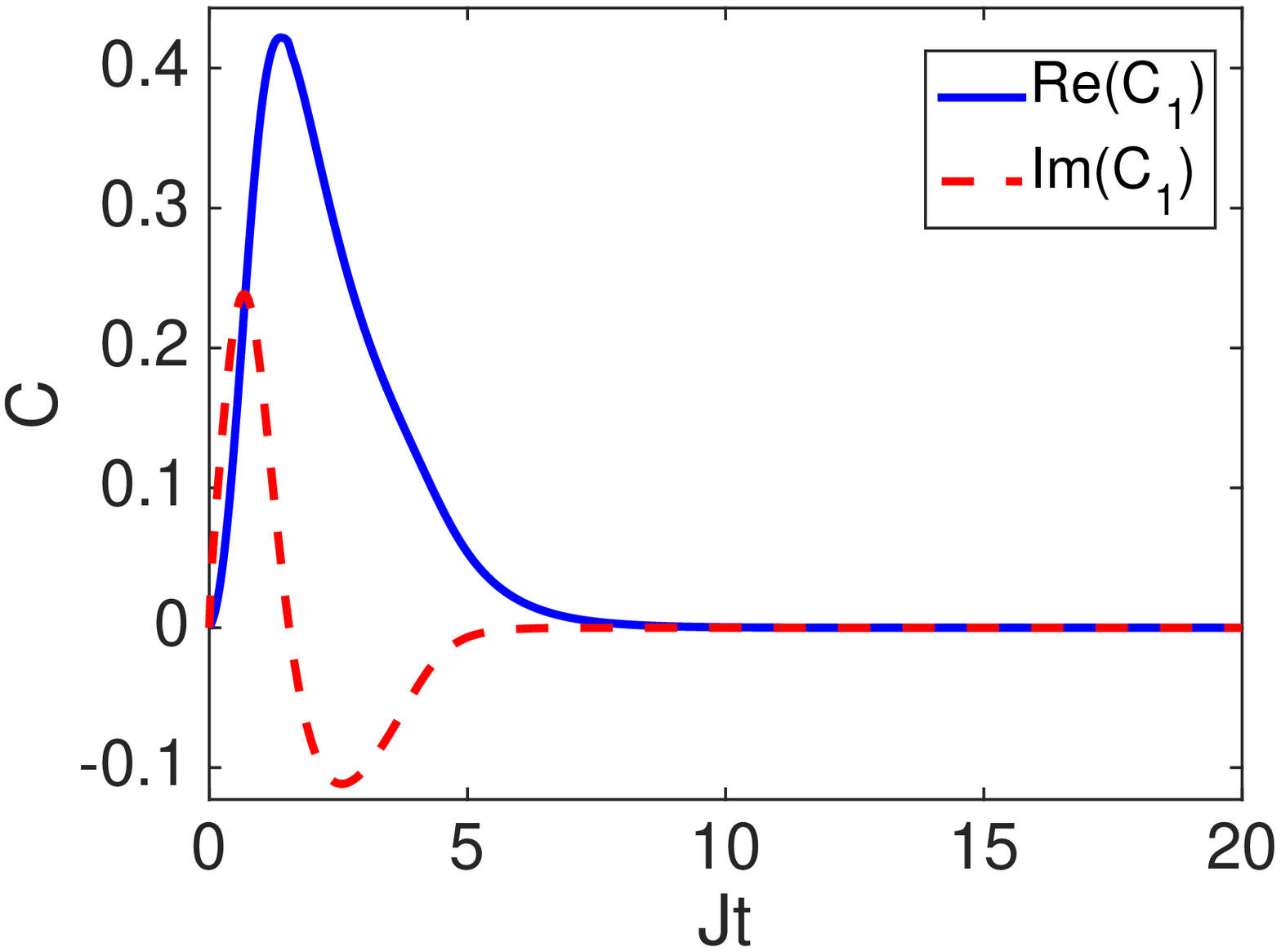}
\includegraphics[width=0.63\columnwidth]{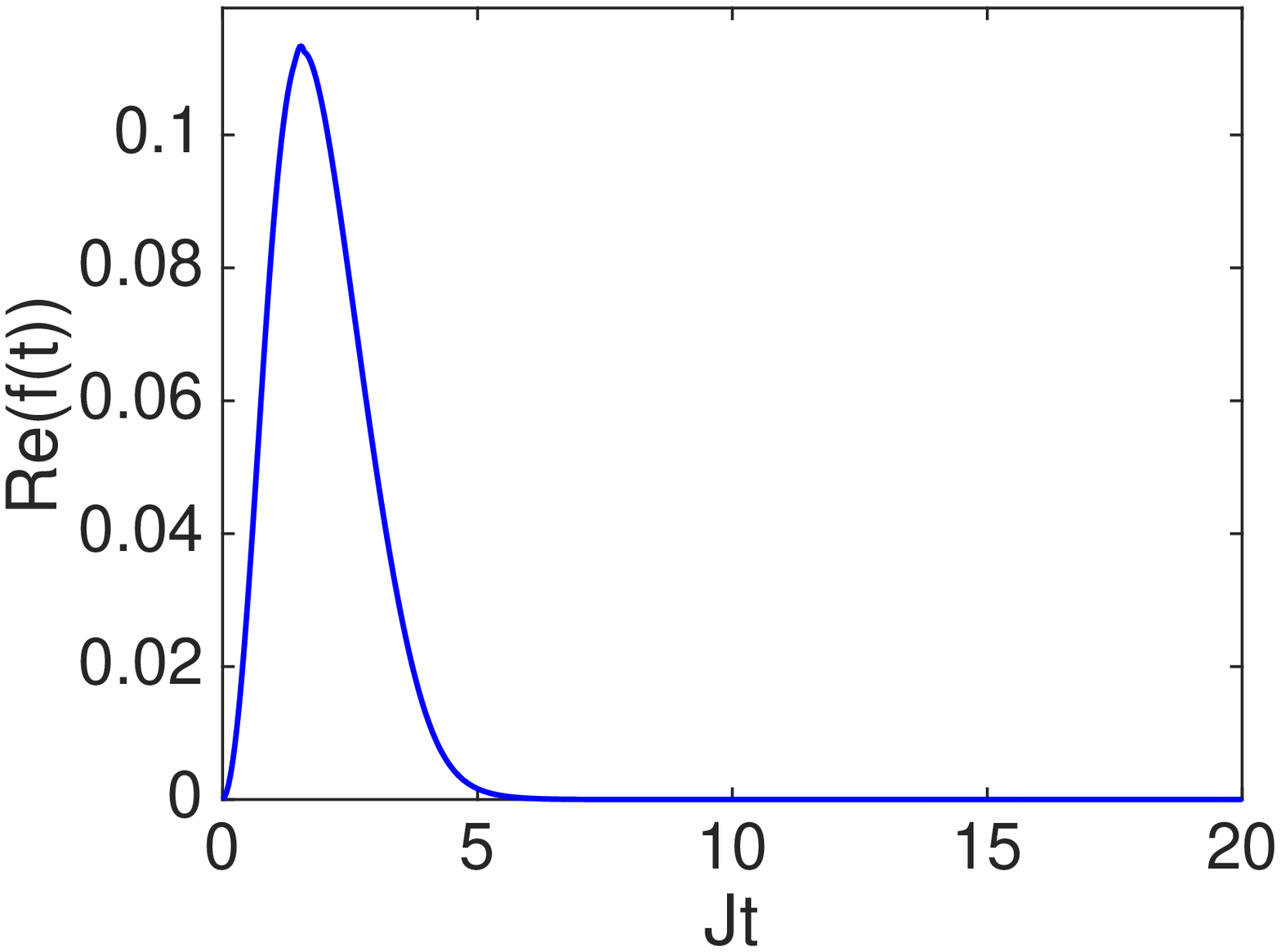}
\includegraphics[width=0.63\columnwidth]{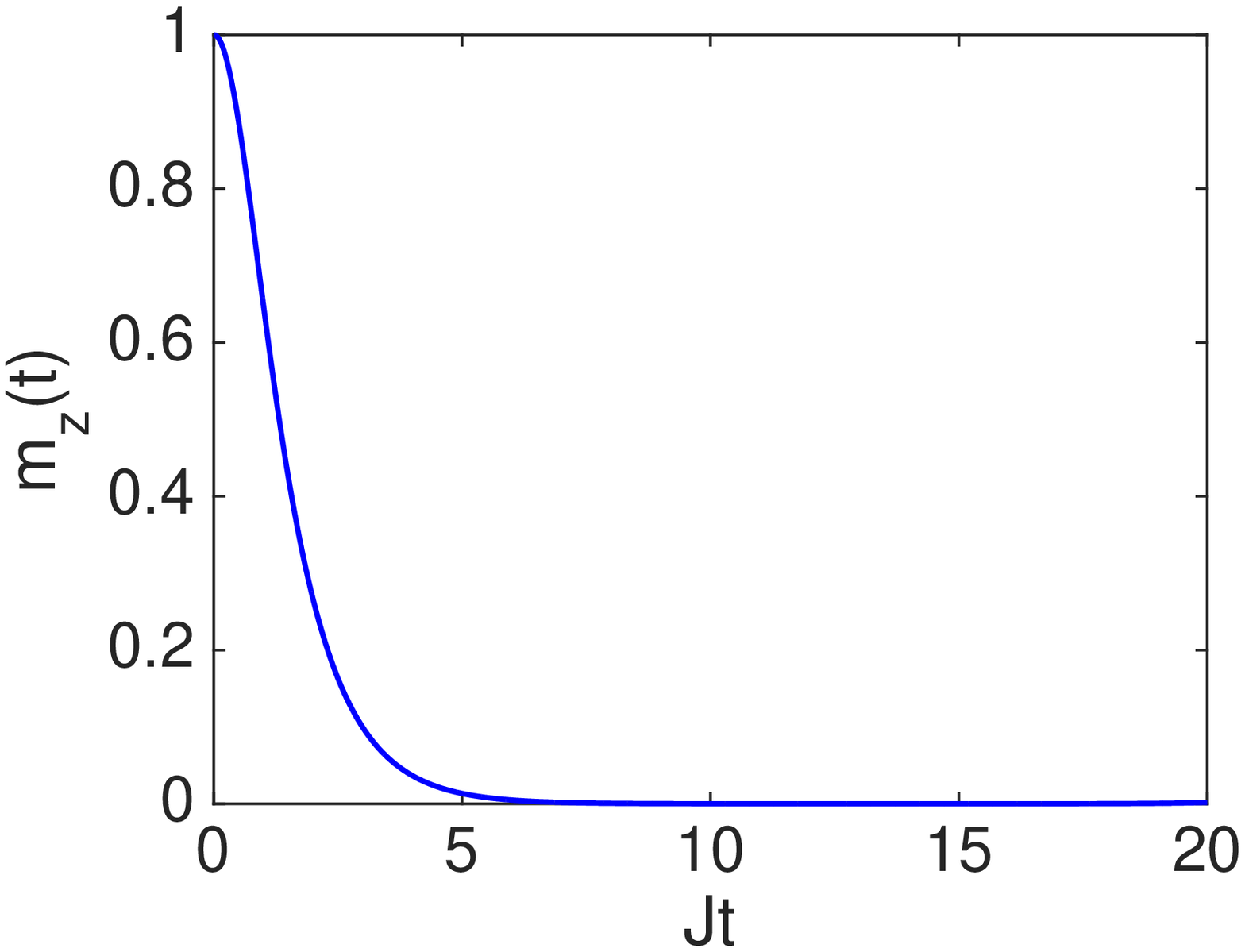}
\caption{Plot of the initial conditions $C_1$ (Left) and 
the corresponding rate function $f_1$ (Center) at 
$\Gamma_{\rm i}=0$ and $h=0+$ with $\Gamma_{\rm f}/J=\Gamma_{\rm d}/J=1/2$. 
The dynamics on the separatrix is not periodic, which is demonstrated 
by the right panel plotting the magnetization 
$m_z(t)=\bra{\Psi(t)}\sigma^z \ket{\Psi(t)}$ computed 
by the semiclassical method in Ref.~\cite{Sciolla:11}.}
\Lfig{separatrix}
\end{center}
\end{figure*}
This figure shows that the rate function asymptotically vanishes as $t\to \infty$, but this does not necessarily imply $|\mathcal{L}(t)|\to 1$ as pointed out at the end of \Rsec{A solvable case:}.

\subsection{Comparison with numerical experiments}
\Lsec{Comparison with numerical}

To validate our semiclassical computations, we here show
the results of numerical experiments and compare them
with the semiclassical results for several parameters.
Our Hamiltonian \NReq{Hamiltonian} commutes with the squared total 
spin operator $\hat{\V{S}}^2=\hat{S}_x^2+\hat{S}_y^2+\hat{S}_z^2$.
For both the quenches from $\Gamma_{\rm i}=0$ and $\infty$, 
the initial state is in the subspace of the total spin $S=N/2$. 
Hence the state of our system preserves this total spin and 
we may consider the time-dependent state inside this subspace.
In the basis of the eigenvalues of $\hat{S}_z$, 
our Hamiltonian is represented in a tridiagonal matrix form and 
we can easily evaluate the time evolution of the state
$\ket{\Psi(t)}=e^{-it\hat{\mathcal{H}}}\ket{\Omega_{\rm b}}$ 
by the LU decomposition. 
The dimension of the subspace is $N+1$ and 
we can treat fairly large size systems. 
However, the computation requires us to take a lot of sums of
complex numbers and the numerical precision tends to be degraded 
as $N$ becomes large. 
This computational difficulty sensitively depends on the parameters 
and below the simulated system sizes are adaptively changed for this reason. 

Figure \NRfig{comp-frominf} is the plots of the rate functions for the quench from $\Gamma_{\rm i}=\infty$.
\begin{figure}[tbp]
\begin{center}
\includegraphics[width=0.47\columnwidth]{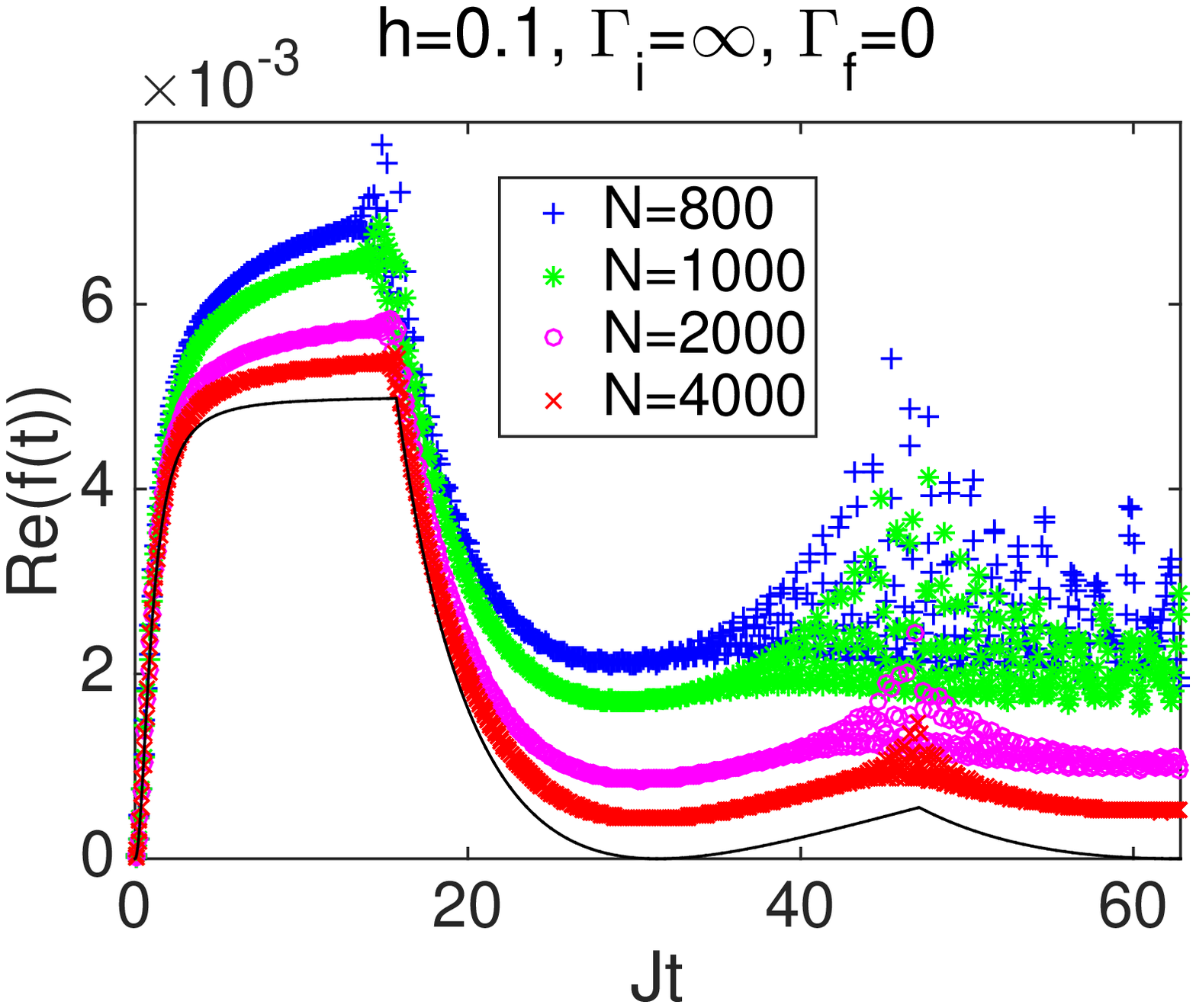}
\includegraphics[width=0.47\columnwidth]{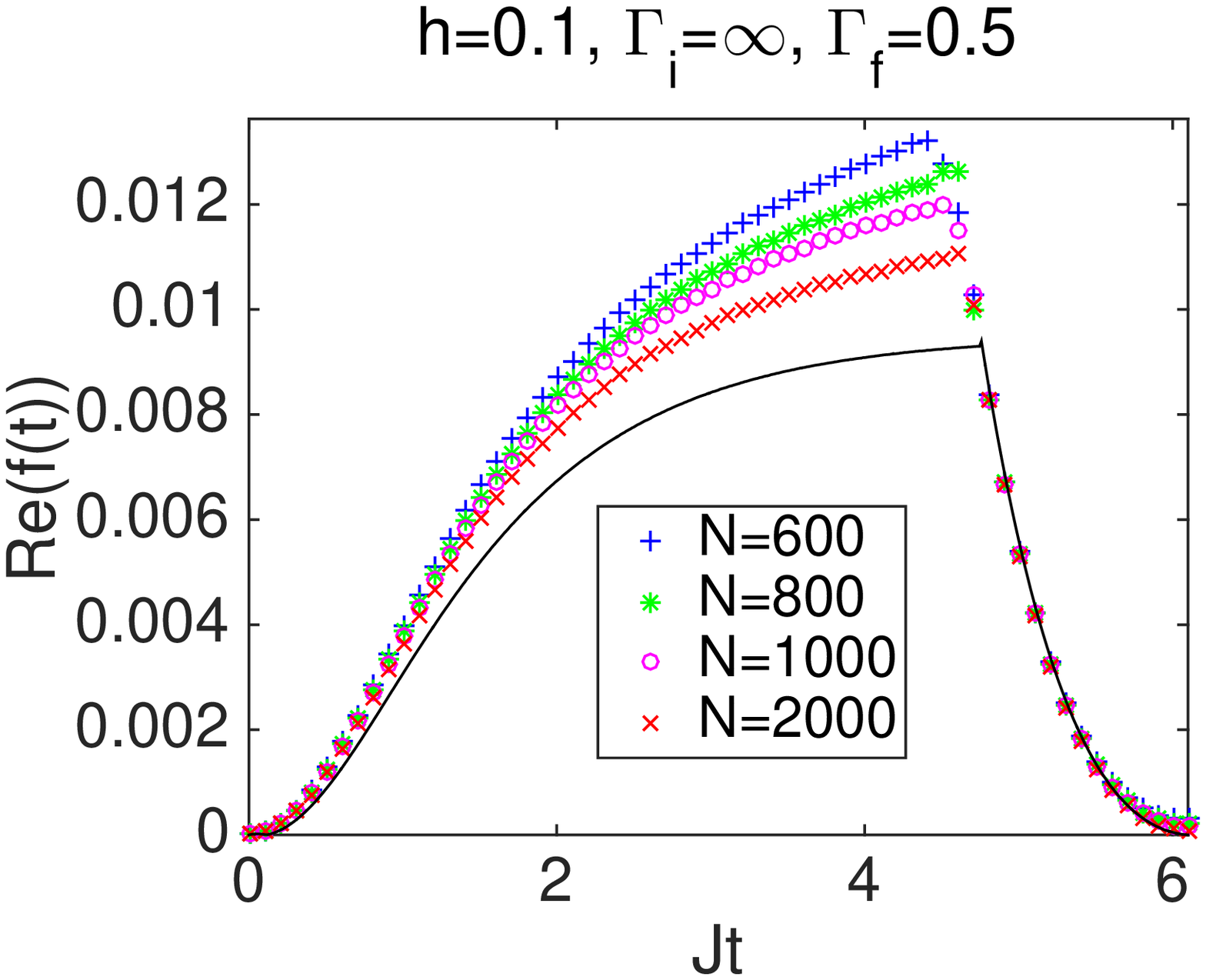}
\includegraphics[width=0.47\columnwidth]{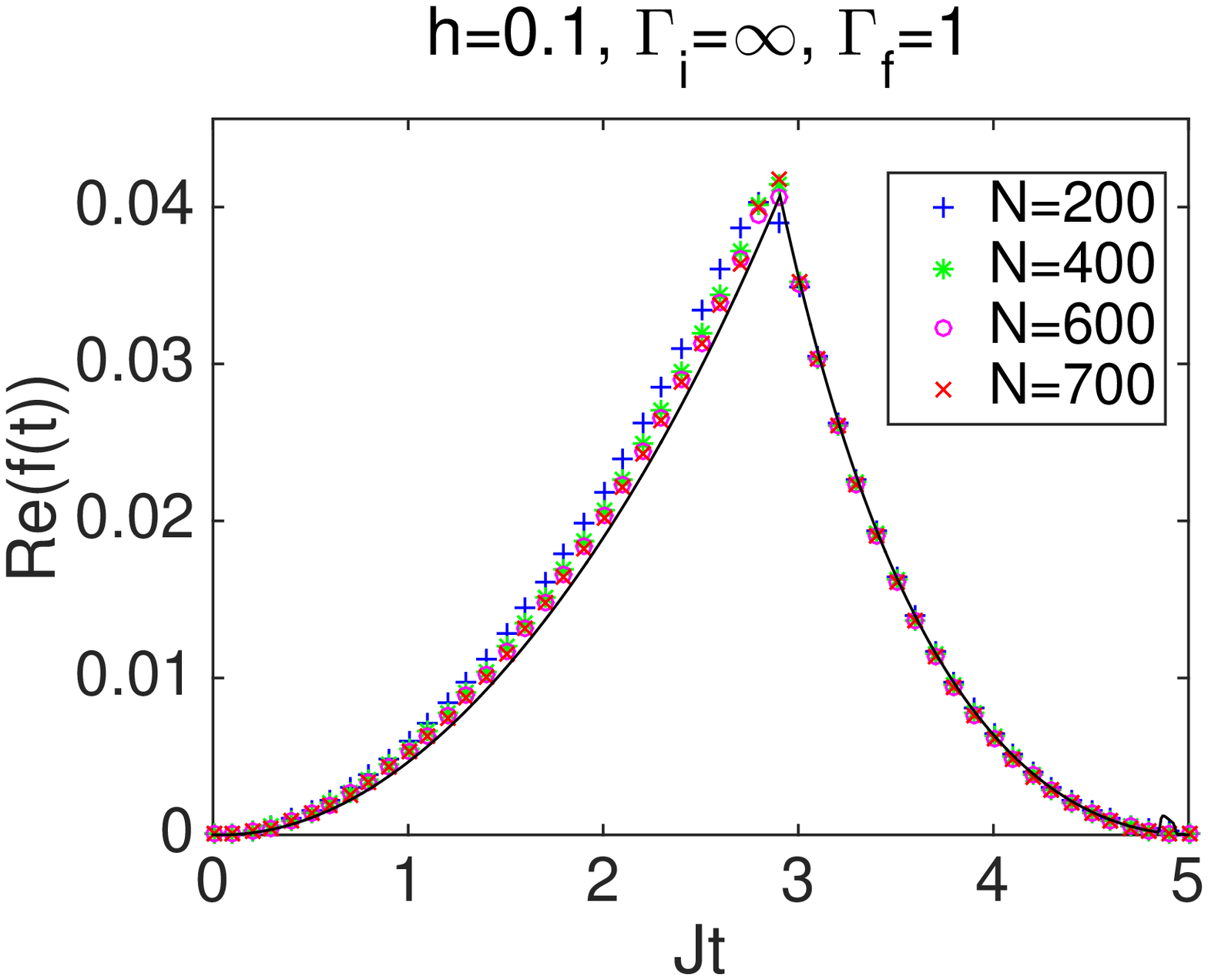}
\includegraphics[width=0.47\columnwidth]{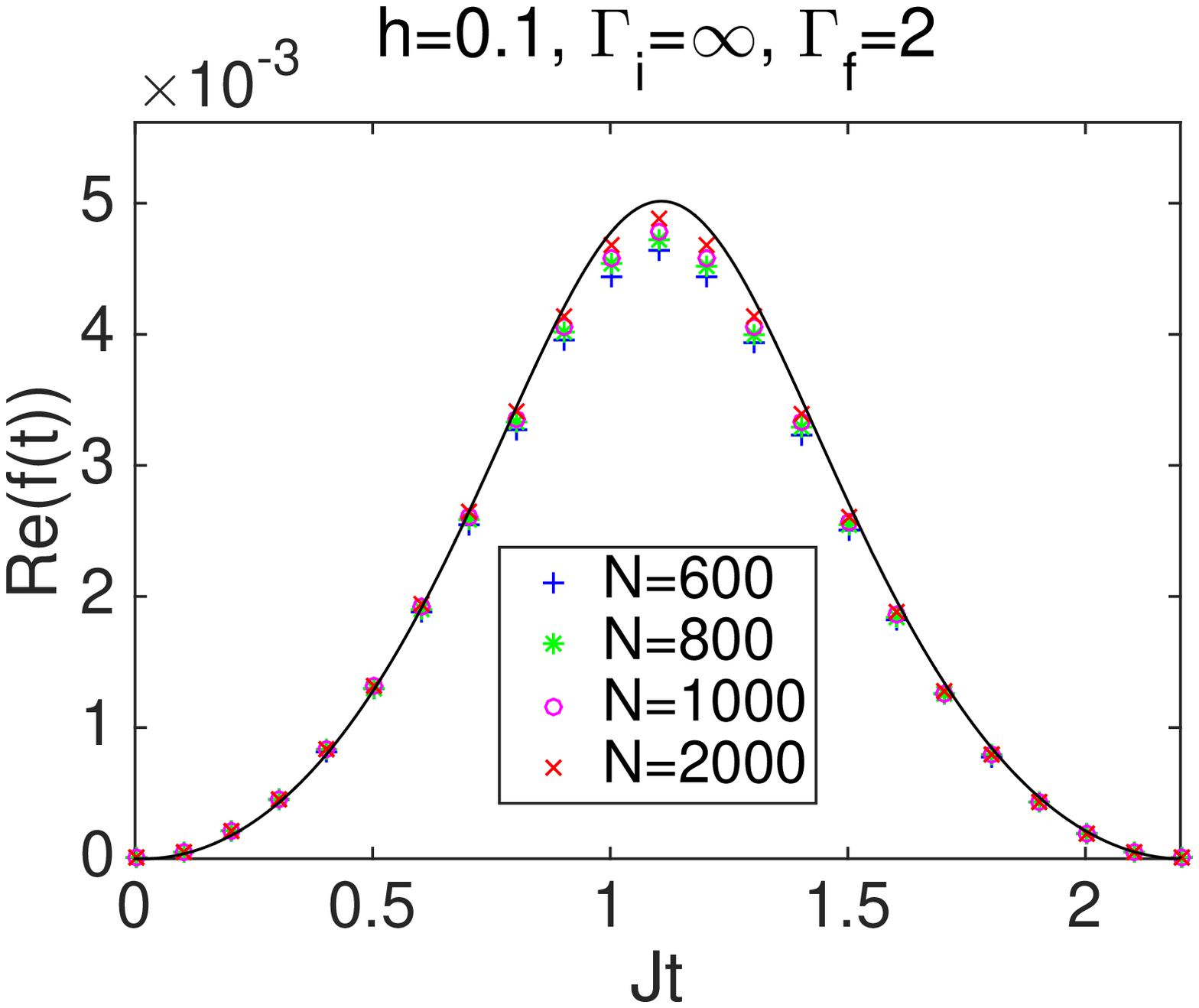}
\caption{The real part of the rate function for 
$\Gamma_{\rm f}/J=0$ (Upper left), $0.5$ (Upper right), 
$1.0$ (Lower left), and $2.0$ (Lower right) 
at $h/J=0.1$ for the quench from $\Gamma_{\rm i}=\infty$. 
The three panels except for the lower right one show the DQPT, 
which is in agreement with the semiclassical computation given by the black solid line. }
\Lfig{comp-frominf}
\end{center}
\end{figure}
The results of numerical simulation show a good agreement with the theoretical
curve denoted by the solid black line, both below and above the
transition point $\Gamma_{\rm fc}(h/J=0.1)/J\approx 1.53$.
This justifies our semiclassical computation.
The upper left panel in \Rfig{comp-frominf} for 
$\Gamma_{\rm f}=0$ and $h/J = 0.1$ 
is compared with the result in \Rfig{Gf0-h01} where the period is given by 
$J\tau=\frac{\pi J}{h} \approx 31.4$. 
We see the consistent agreement between the numerical and semiclassical computations.
The deviation for the whole time and the oscillating behavior at
large $t$ are considered to be due to the finite size effect. 

Figure \NRfig{comp-fromzero} represents the result of 
a quench from $\Gamma_{\rm i}=0$ at $h=0+$.
\begin{figure}[tbp]
\begin{center}
\includegraphics[width=0.47\columnwidth]{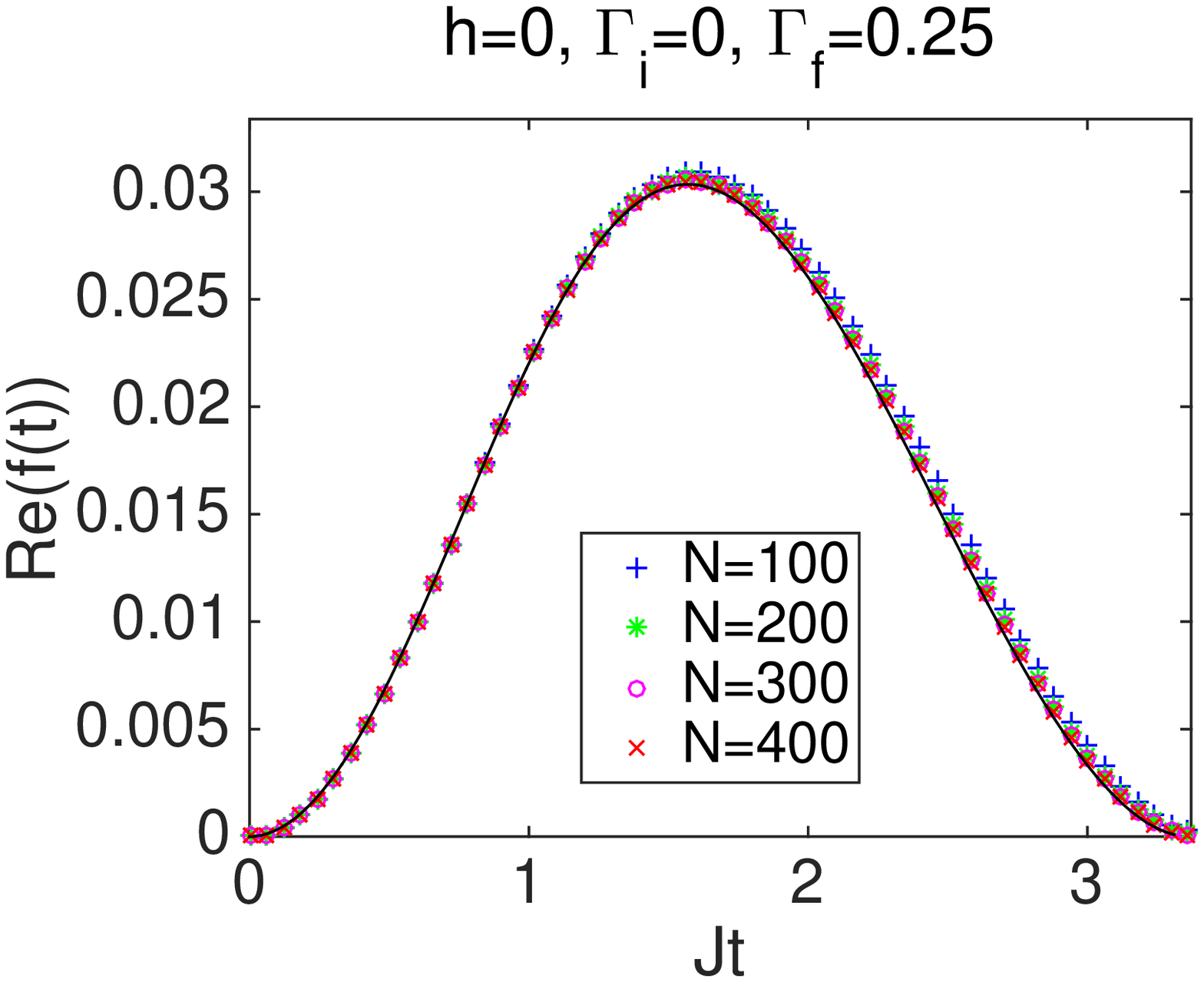}
\includegraphics[width=0.47\columnwidth]{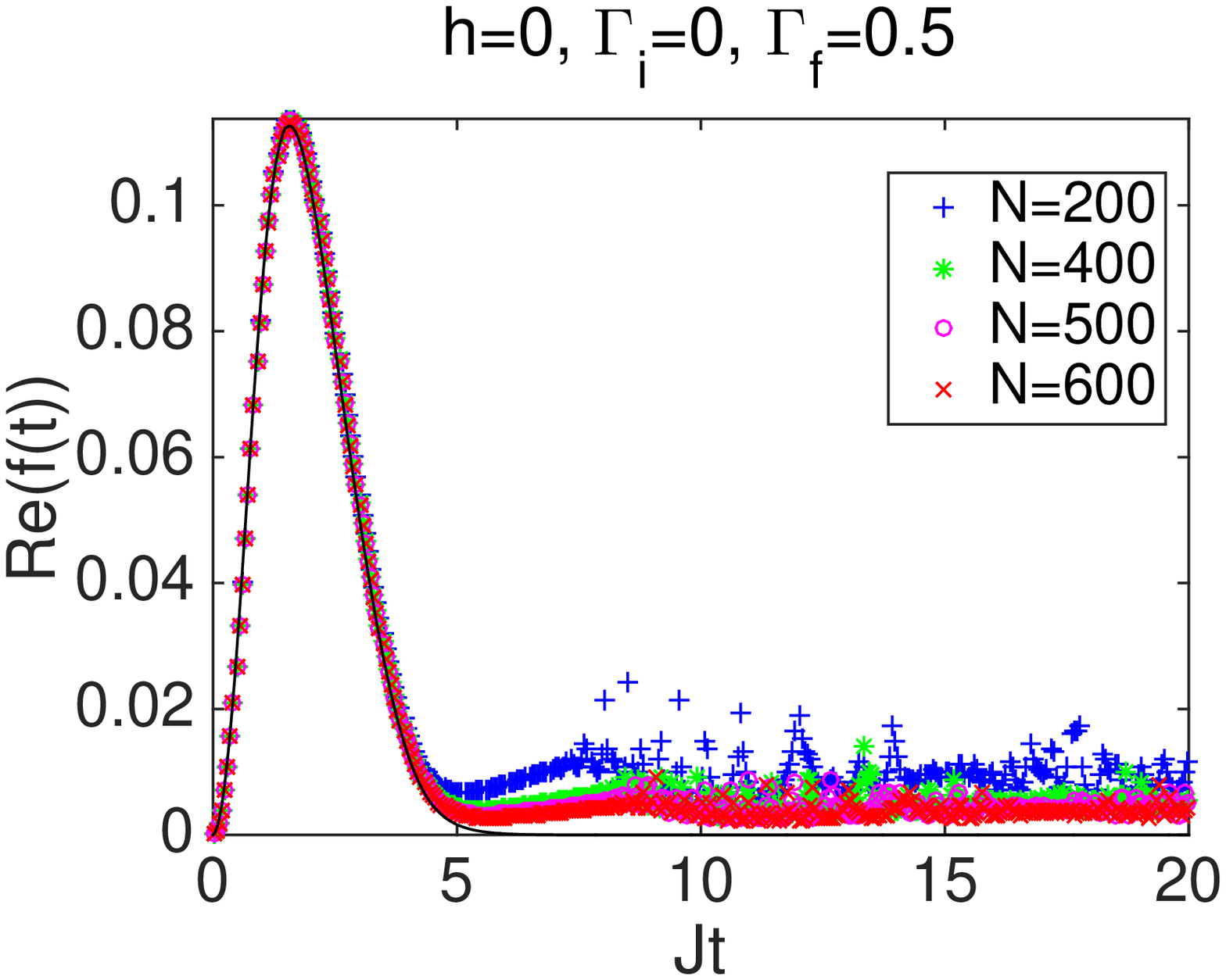}
\includegraphics[width=0.47\columnwidth]{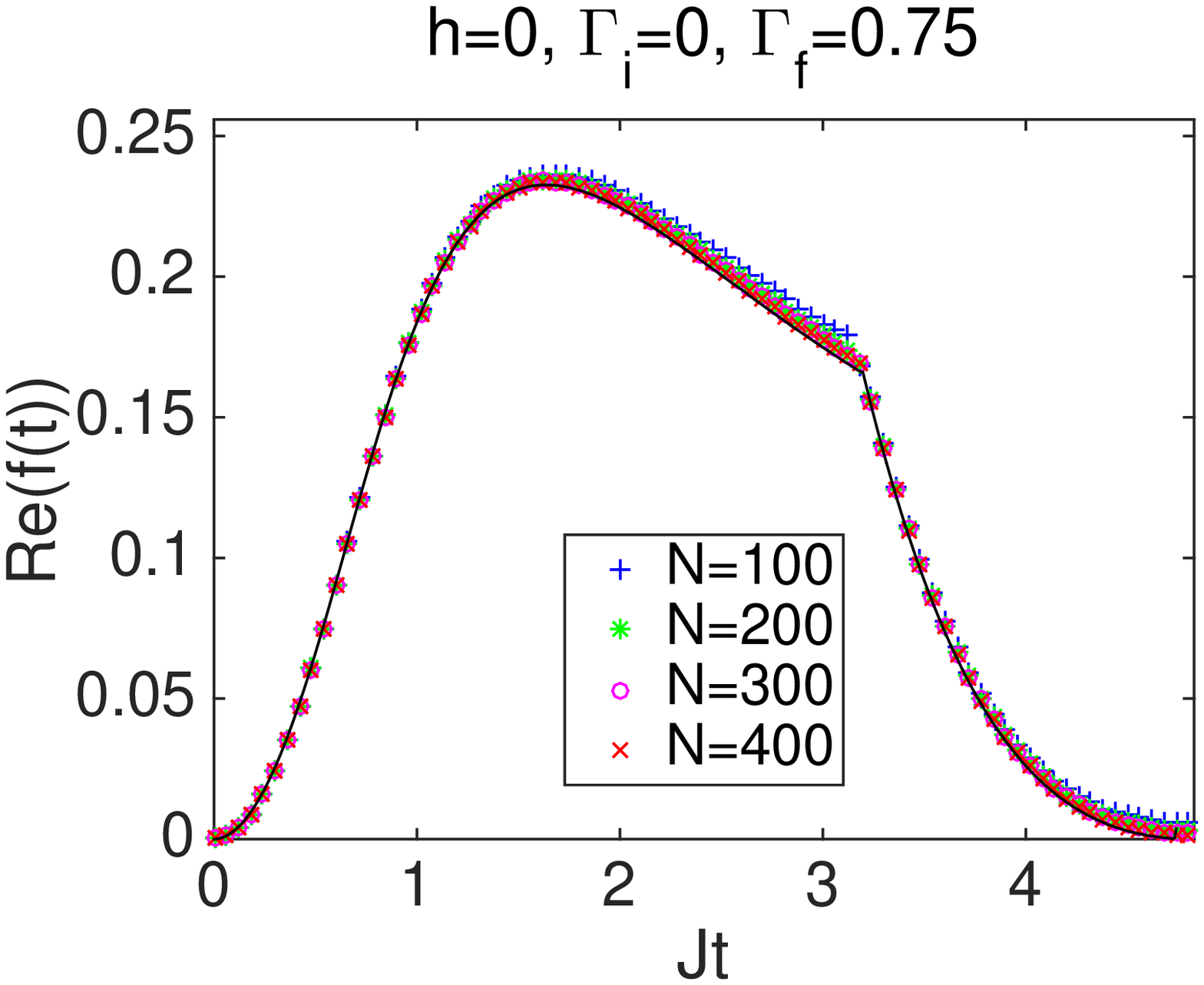}
\includegraphics[width=0.47\columnwidth]{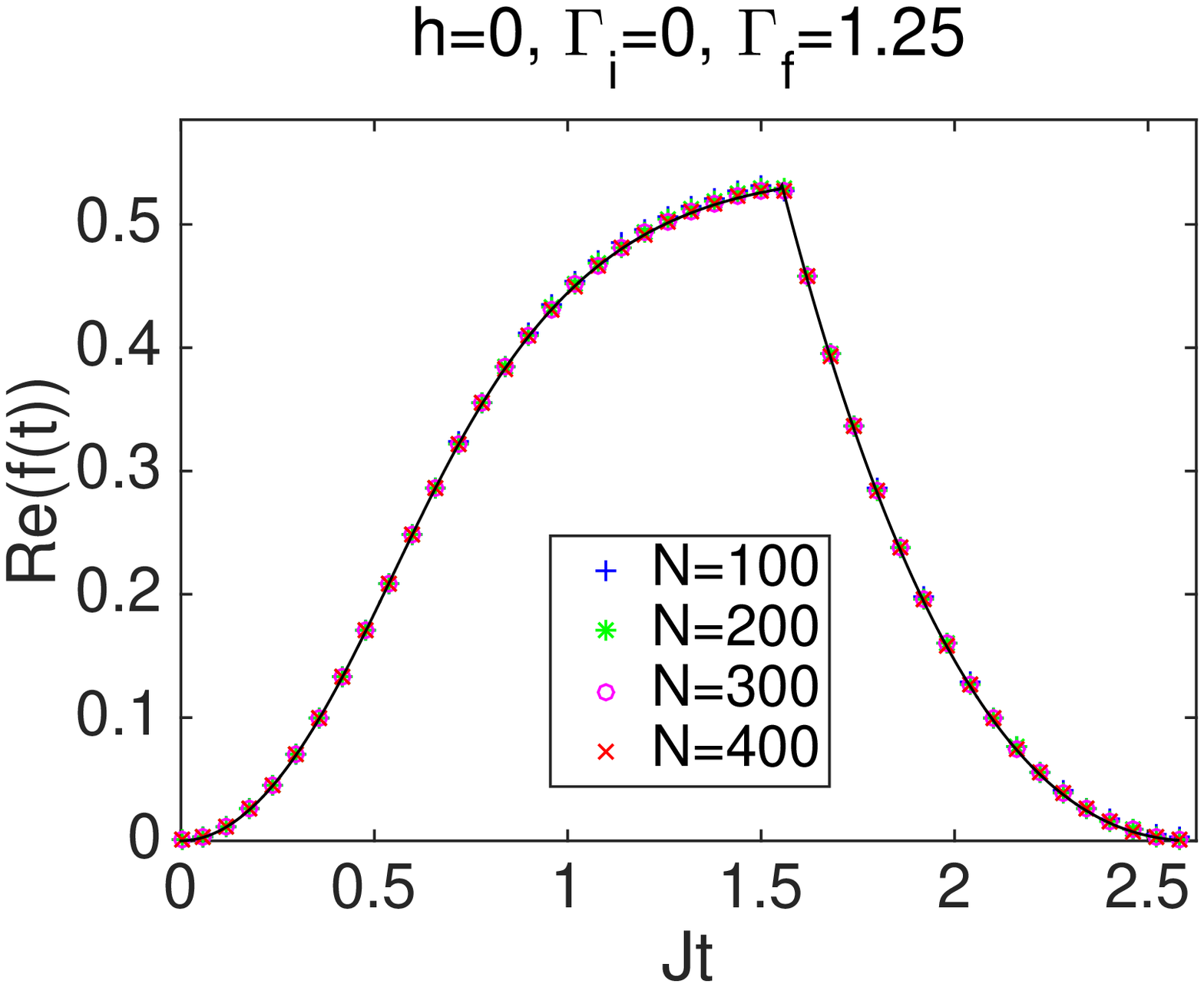}
\caption{The real part of the rate function 
for $\Gamma_{\rm f}/J=0.25$ (Upper left), $0.5$ (Upper right), 
$0.75$ (Lower left), and $1.25$ (Lower right) at $h=0+$ 
for the quench from  $\Gamma_{\rm i}=0$. 
The upper two panels do not show any DQPT while the lower ones do. 
The upper right panel is for the separatrix and the corresponding rate function 
shows a monotonic decay after a smooth peak.}
\Lfig{comp-fromzero}
\end{center}
\end{figure}
Again, the numerical results show a fairly good agreement with the semiclassical curve. At the separatrix, $\Gamma_{\rm f}/J=1/2$, the monotonic decay of the rate function after a single peak is well reproduced by the numerics, validating our semiclassical computation even at a special point of the dynamics.  

\section{Discussion and summary}
\label{Discussion}

In this paper, we have invented a computational method for 
the Loschmidt amplitude based on the complex semiclassical approach, 
and applied it to the transverse field Ising model 
with a symmetry breaking field in the infinite dimension. 
Two quantum quenches, from zero and infinite transverse fields, 
have been examined. 
From the behavior of the rate function, the presence or absence 
of the DQPTs have been captured.
The phase diagrams have been mapped out in the plane of
final transverse field and symmetry breaking field. 
These results have been examined by numerical simulations independently that
solve the Schr\"odinger equation literally, 
which fully supports our semiclassical computations.

Although our computational method has succeeded in unveiling several
properties of the Loschmidt amplitude, its physical implications
are still unclear. \v{Z}unkovi\v{c} \textit{et al.} have pointed a
connection
between the Loschmidt amplitude and the order parameter in the steady state long after the quench
\cite{Zunkovic:16}. However we have not
found such a connection as far as the quench from $\Gamma_{\rm i} =
\infty$ is concerned.
Therefore the presented result might add a further mystery on the DQPT. 
Disentangling DQPTs of the Loschmidt amplitude and an order parameter
may open a new comprehension on quantum dynamics.  

An experimental observation of a DQPT is a fascinating topic. A very recent work~\cite{Flaschner:16} has actually observed DQPTs using a certain topological nature of the singularity~\cite{Vajna:15,Budich:16,Sharma:16}. Unfortunately, this is possible only in non-interacting systems and its generalization to interacting systems is unclear. Although there are some other experiments~\cite{Zhang:09,Zangara:16} observing the Loschmidt amplitude, their methods rely on the smallness of the system or certain locality of the phenomena. The application of their methods to global phenomena in many-spin systems is again nontrivial. Our model, the Ising model with long range interactions, itself can be realized in a trapped ion system~\cite{Islam:13}. Another recent experiment on this system has observed nontrivial cusps in the probability to return to the ground-state manifold, giving a clear evidence of the DQPT~\cite{Heyl:14,Jurcevic:16}. Their setup corresponds to $\Gamma_i=0$ and $h=0+$ in the present paper, and we expect that further nontrivial results can be obtained in other setups according to our findings. Such additional experiments are encouraged.

A more direct application of our method might be found in quantum
engineering or computing. 
In those disciplines, it is an important problem to estimate 
the probability achieving a desired state in
certain quantum processes. 
For example in quantum annealing~\cite{QA1,QA2,QA3}, 
the probability to find the ground state is an important object 
to be calculated. 
Using techniques from the spin glass theory~\cite{Obuchi:12,Takahashi:13} 
combined with the present method, its typical value might be evaluated. 
This will provide a theoretical challenge for both quantum mechanics and random spin systems.   

\section*{Acknowledgements} 

The authors thank Makoto Negoro for discussions about experimental relevance. This work was supported by KAKENHI Nos. 26870185 (TO), 26400402 (SS), and 26400385 (KT). 



\begin{thebibliography}{99}

\bibitem{Polkovnikov:11} 
A. Polkovnikov, K. Sengupta, A. Silva, and M. Vengalattore, 
Nonequilibrium dynamics of closed interacting quantum systems, 
Rev. Mod. Phys. \textbf{83}, 863 (2011).

\bibitem{Heyl:13} 
M. Heyl, A. Polkovnikov, and S. Kehrein, 
Dynamical Quantum Phase Transitions in the Transverse-Field Ising Model, 
Phys. Rev. Lett. \textbf{110}, 135704 (2013).

\bibitem{Zunkovic:16} 
B. \v{Z}unkovi\v{c}, M. Heyl, M. Knap, and A. Silva, 
Dynamical quantum phase transitions in spin chains with long-range interactions: 
Merging different concepts of non-equilibrium criticality, 
arXiv:1609.08482.

\bibitem{Heyl:14} 
M. Heyl, Dynamical Quantum Phase Transitions in Systems with Broken-Symmetry Phases, 
Phys. Rev. Lett. \textbf{113}, 205701 (2014).

\bibitem{Heyl:15} 
M. Heyl, Scaling and Universality at Dynamical Quantum Phase Transitions, 
Phys. Rev. Lett. \textbf{115}, 140602 (2015).

\bibitem{Zunkovic:15} 
B. \v{Z}unkovi\v{c}, A. Silva, and M. Fabrizio, 
Dynamical phase transitions and Loschmidt echo in the infinite-range XY model, 
Phil. Trans. R. Soc. A \textbf{374}, 20150160 (2015).

\bibitem{Karrasch:13}
C. Karrasch and D. Schuricht, 
Dynamical phase transitions after quenches in nonintegrable models, 
Phys. Rev. B \textbf{87}, 195104 (2013).

\bibitem{Vajna:14} 
S. Vajna and B. D\'ora, 
Disentangling dynamical phase transitions from equilibrium phase transitions, 
Phys. Rev. B \textbf{89}, 161105(R) (2014).

\bibitem{Sharma:15} 
S. Sharma, S. Suzuki, and A. Dutta, 
Quenches and dynamical phase transitions in a nonintegrable quantum Ising model, 
Phys. Rev. B \textbf{92}, 104306 (2015).

\bibitem{Divakaran:16}
U. Divakaran, S. Sharma, and A. Dutta,
Tuning the presence of dynamical phase transitions in a generalized \textit{XY} spin chain, Phys. Rev. E \textbf{93}, 052133 (2016).

\bibitem{Gambassi:11-1}
A. Gambassi and A. Silva, 
Statics of the work in quantum quenches, universality and the critical Casimir effect, 
arXiv:1106.2671.

\bibitem{Gambassi:11-2} 
A. Gambassi and P. Calabrese, Quantum quenches as classical critical films, EPL {\bf 95}, 66007  (2011).
\bibitem{Chiocchetta:15} 
A. Chiocchetta, M. Tavora, A. Gambassi, and A. Mitra, Short-time universal scaling in an isolated quantum system after a quench, Phys. Rev. B {\bf 91}, 220302 (2015).
\bibitem{Chiocchetta:16} 
A. Chiocchetta, M. Tavora, A. Gambassi, and A. Mitra, Short-time universal scaling and light-cone dynamics after a quench in an isolated quantum system in $d$ spatial dimensions, Phys. Rev. B {\bf 94}, 134311 (2016).
\bibitem{Maraga:15}
A. Maraga, A. Chiocchetta, A. Mitra, and A. Gambassi, Aging and coarsening in isolated quantum systems after a quench: Exact results for the quantum $O(N)$ model with $N\to \infty$, Phys. Rev. E {\bf 92}, 042151 (2015).


\bibitem{Smacchia:13}
P. Smacchia and A. Silva, 
Work distribution and edge singularities for generic time-dependent protocols
in extended systems, 
Phys. Rev. E \textbf{88}, 042109 (2013).

\bibitem{Gambassi:12}
A. Gambassi and A. Silva, 
Large Deviations and Universality in Quantum Quenches, 
Phys. Rev. Lett. \textbf{109}, 250602 (2012).

\bibitem{NishimoriBook} 
H. Nishimori and G. Ortiz, 
\textit{Elements of Phase Transitions and Critical Phenomena} 
(Oxford University Press, Oxford, 2011).

\bibitem{Jurcevic:16}
P. Jurcevic, H. Shen, P. Hauke, C. Maier, T. Brydges, C. Hempel,
B. P. Lanyon, M. Heyl, R. Blatt, and C. F. Roos,
Direct observation of dynamical quantum phase transitions in an
interacting many-body system, arXiv:1608.05616.

\bibitem{SuzukiBook} 
S. Suzuki, J-i. Inoue, and B. K. Chakrabarti,
\textit{Quantum Ising Phases and Transitions in Transverse Ising Models}, 
Lecture Notes in Physics Vol. 862 (Springer, Berlin, 2013).

\bibitem{Sciolla:10} 
B. Sciolla and G. Biroli, 
Quantum Quenches and Off-Equilibrium Dynamical Transition 
in the Infinite-Dimensional Bose-Hubbard Model, 
Phys. Rev. Lett. \textbf{105}, 220401 (2010).

\bibitem{Sciolla:11} 
B. Sciolla and G. Biroli, 
Dynamical transitions and quantum quenches in mean-field models, 
J. Stat. Mech. P11003 (2011).

\bibitem{Klauder:79} 
J. R. Klauder,
Path integrals and stationary-phase approximations, 
Phys. Rev. D {\bf 19}, 2349 (1979).

\bibitem{Alscher:99} 
A. Alscher and  H. Grabert,  
Semiclassical dynamics of a spin-1/2 in an arbitrary magnetic field, 
J. Phys. A: Math. Gen. {\bf 32}, 4907 (1999).

\bibitem{Obuchi:12} 
T. Obuchi and K. Takahashi, 
Dynamical singularities of glassy systems in a quantum quench, 
Phys. Rev. E {\bf 86}, 051125 (2012).

\bibitem{Takahashi:13} 
K. Takahashi and T. Obuchi, 
Zeros of the partition function and dynamical singularities in spin-glass systems, 
J. Phys.: Conf. Ser. {\bf 473}, 012023 (2013).

\bibitem{Flaschner:16} 
N. Fl\"aschner, D. Vogel, M. Tarnowski, 
B. S. Rem. D.-S. L\"uhmann, M. Heyl, J. C. Budich, L. Mathey, 
K. Sengstock, and C. Weitenberg, 
Observation of a dynamical topological phase transition, 
arXiv:1608.05616.

\bibitem{Vajna:15}
S. Vajna and B. D\'ora, 
Topological classification of dynamical phase transitions, 
Phys. Rev. B \textbf{91}, 155127 (2015).

 \bibitem{Budich:16} 
J. C. Budich and M. Heyl, 
Dynamical topological order parameters far from equilibrium, 
Phys. Rev. B \textbf{93}, 085416 (2016).

\bibitem{Sharma:16}
S. Sharma, U. Divakaran, A. Polkovnikov, and A. Dutta, 
Slow quenches in a quantum Ising chain: Dynamical phase transitions and topology, 
Phys. Rev. B \textbf{93}, 144306 (2016).

\bibitem{Zhang:09}
J. Zhang, F. M. Cucchietti, C. M. Chandrashekar, M. Laforest, 
C. A. Ryan, M. Ditty, A. Hubbard, J. K. Gamble, and R. Laflamme, 
Direct observation of quantum criticality in Ising spin chains, 
Phys. Rev. A {\bf 79}, 012305 (2009).

\bibitem{Zangara:16} 
P. R. Zangara, D. Bendersky, P. R. Levstein, and H. M. Pastawski, 
Loschmidt echo in many-spin systems: 
contrasting time scales of local and global measurements, 
Phil. Trans. R. Soc. A. {\bf 374}, 20150163 (2016).

\bibitem{Islam:13} 
R. Islam, C. Senko, W. C. Campbell, S. Korenblit, J. Smith, 
A. Lee, E. E. Edwards, C.-C. J. Wang, J. K. Freericks, and C. Monroe, 
Emergence and frustration of magnetism with variable-range interactions 
in a quantum simulator, 
Science \textbf{340}, 583 (2013).
	
\bibitem{QA1} 
A. Das and B. K. Chakrabarti, 
\textit{Quantum Annealing and Related Optimization Methods} (Springer, 2005).

\bibitem{QA2} 
D. de Falco and D. Tamascelli, 
An introduction to quantum annealing, 
RAIRO-Theor. Inf. Appl. {\bf 45}, 99 (2011).

\bibitem{QA3} 
S. Morita and H. Nishimori, 
Mathematical foundation of quantum annealing, 
J. Math. Phys. {\bf 49}, 125210 (2008).

\end{thebibliography}
\end{document}